%% file: main.tex
\documentclass{article} % For LaTeX2e
\usepackage{iclr2027_conference,times}
\input{math_commands.tex}

\usepackage{graphicx}
\usepackage[pagebackref=true,breaklinks=true,colorlinks,bookmarks=false,urlcolor=blue,citecolor=blue,linkcolor=blue]{hyperref}
\usepackage{url}
\usepackage{booktabs}
\usepackage[table]{xcolor} % 提供 \rowcolor
\definecolor{Gray}{gray}{0.9} % 定义 Gray（0=黑,1=白，可改深浅）

\usepackage{pifont} % 导入 pifont 宏包
\usepackage{multirow} 

\newcommand{\ourframework}{\textsc{LensBridge}}
\newcommand{\red}[1]{{\color{red}#1}}
\newcommand{\blue}[1]{{\color{blue}#1}}
\usepackage{pgfplots}
\usepackage{tikz}
\usepackage{amsmath}

\pgfplotsset{compat=1.18}
\title{LensBridge: Frequency-Guided Compound Degradation Adaptation for Lens Aberration Correction and Veiling Glare Removal}

\makeatletter
\renewcommand*{\@fnsymbol}[1]{\ensuremath{\ifcase#1\or *\or \dagger\or \ddagger\or
    \mathsection\or \mathparagraph\or \|\or **\or \dagger\dagger
    \or \ddagger\ddagger \else\@ctrerr\fi}}
\makeatother

\author{Xiaolong Qian$^{1}$ \quad
Zhonghua Yi$^{1}$ \quad
Qi Jiang$^{1}$ \quad
Kailun Yang$^{2}$ \quad
Shuhang Xie$^{1}$\\
\textbf{Shaohua Gao}$^{1}$ \quad
\textbf{Kaiwei Wang}$^{1,}$\thanks{Corresponding author.}\\
$^1$Zhejiang University \quad
$^2$Hunan University
}

\usepackage{caption}
\usepackage{wrapfig}
\usepackage{enumitem}
\usepackage{paralist}
\usepackage{adjustbox}

\newcommand{\methodhead}[1]{%
    \begin{tabular}[c]{@{}c@{}}#1\end{tabular}%
}
\newcommand{\best}[1]{\textcolor{red}{\textbf{#1}}}
\newcommand{\second}[1]{\textcolor{blue}{\underline{#1}}}

\definecolor{AberrationColor}{HTML}{946600}
\definecolor{CleanColor}{HTML}{6B6B6B}
\definecolor{VGColor}{HTML}{785488}

\newcommand{\AB}[1]{\textcolor{AberrationColor}{#1}}

\newcommand{\VG}[1]{\textcolor{VGColor}{#1}}

\usepackage{bm}
\usepackage{pifont}
\iclrfinalcopy %
\begin{document}

\maketitle

\begin{abstract}
Simplified optical systems often exhibit residual lens aberrations and Veiling Glare (VG), resulting in spatially varying blur and contrast reduction. Large-scale Lens Libraries (LensLib) enable reusable aberration correction models by covering diverse Point Spread Functions (PSFs), but their aberration-only training distribution does not include target-specific veiling glare. Extending such foundations to compound degradation is challenging because realistic target-system compound pairs are difficult to obtain. To address this challenge, we propose \ourframework{}, a two-stage framework that first establishes a reusable aberration correction foundation and then adapts it to compound optical degradation using only a few unpaired target observations. In Stage I, we build a PSF-aware one-step diffusion foundation by constructing discrete degradation priors from LensLib PSFs and learning to retrieve them directly from aberrated images, enabling PSF-aware correction without requiring explicit PSF at inference. In Stage II, we adapt this foundation to compound degradation through frequency-domain guidance. Our analysis suggests that aberrations affect a broad frequency range, whereas additional veiling glare exhibits stronger low-frequency amplitude variations, motivating a shared target-domain Fourier-amplitude statistic. At the data level, Frequency-guided Degradation Completion (FDC) transfers target low-frequency characteristics to LensLib aberrated images while preserving aberration structures to synthesize compound training pairs; at the model level, Frequency-guided Pseudo Decomposition (FPD) forms aberration- and VG-dominant pseudo observations to condition separate adaptation branches. Extensive experiments across multiple optical systems demonstrate that \ourframework{} effectively extends reusable aberration correction foundations to joint aberration correction and veiling glare removal without target-system paired supervision. All code will be available at \url{https://github.com/XiaolongQian/LensBridge}.
\end{abstract}

\section{Introduction}
\label{sec:intro}
\input{Tex/Intro}

\section{Related Work}
\label{sec:related}
\input{Tex/Related_Work}

\section{Methodology}
\label{sec:method}
\input{Tex/Method}

\section{Experiments}
\label{sec:exp}
\input{Tex/Exp}

\section{Conclusion}
\label{sec:conclusion}
\input{Tex/conclusion}

\clearpage
\subsection*{AI use statement}
In this work, we used generative AI tools to aid and polish the writing of the manuscript.
We have reviewed and revised all AI-assisted text and verified the technical content and citations.
We take responsibility for the final content of this work, including text and claims produced with the aid of generative AI.

\bibliography{iclr2027_conference}
\bibliographystyle{iclr2027_conference}
\clearpage

\appendix

\section{Additional Implementation Details}
\label{sec:implementation_details}
\noindent\textbf{Training Details.}
For the Stage-I foundation, training proceeds in two phases, with each phase optimized for $30$K iterations.
In Phase 1, we jointly optimize the PSF encoder, discrete codebook, attention-based condition mapping module, and aberration-aware LoRA branch using
$\mathcal{L}_{\mathrm{rec}}+\mathcal{L}_{\mathrm{VQ}}$.
This establishes the PSF-derived discrete aberration prior space.
In Phase 2, we freeze the PSF encoder and codebook and optimize the image-based degradation predictor together with the restoration adaptation modules.
The frozen PSF branch provides teacher supervision through $\mathcal{L}_{\mathrm{FM}}$, encouraging the image-based predictor to retrieve consistent entries from the established codebook.
The Phase-2 objective is $\mathcal{L}_{\mathrm{rec}}+\mathcal{L}_{\mathrm{FM}}$.
For target-system adaptation, we initialize from the Stage-I model and adapt each optical system independently using a limited number of unpaired target observations.
Compared with the foundation training, Stage-II requires lightweight optimization with only $3$K iterations for each target system.
The pretrained SD-Turbo weights and discrete codebook remain frozen, while the image-based degradation predictor, attention-based condition mapping modules, VG encoder, and LoRA parameters of both branches are jointly optimized.
The aberration branch is initialized from Stage I, while the VG branch is newly introduced.
FDC-generated compound pairs are jointly sampled with the original LensLib aberration-only pairs during adaptation.
Since the aberration-dominant observations generated by FPD are pseudo observations rather than physically exact aberration-only inputs, the feature matching loss is retained to preserve consistency with the Stage-I aberration prior space.
The Stage-II objective combines the reconstruction loss and feature matching loss, i.e.,
$\mathcal{L}_{\mathrm{II}}=\mathcal{L}_{\mathrm{rec}}+\mathcal{L}_{\mathrm{FM}}$.

\noindent\textbf{Target Adaptation Data.}
Tab.~\ref{tab:adaptation_data} summarizes the number of unpaired target observations used for Stage-II adaptation across different optical systems.
For the SL and MRL systems, both \texttt{Screen} and \texttt{Real} domains are evaluated, while CAYE, Nano, and UDC provide additional real-world cases beyond the main benchmarks.

\input{Table/few_shot_numbers}

\section{Additional Analysis}
\begin{figure*}[h]
  \centering
  %\vspace{-1em}
  \includegraphics[width=0.55\textwidth]{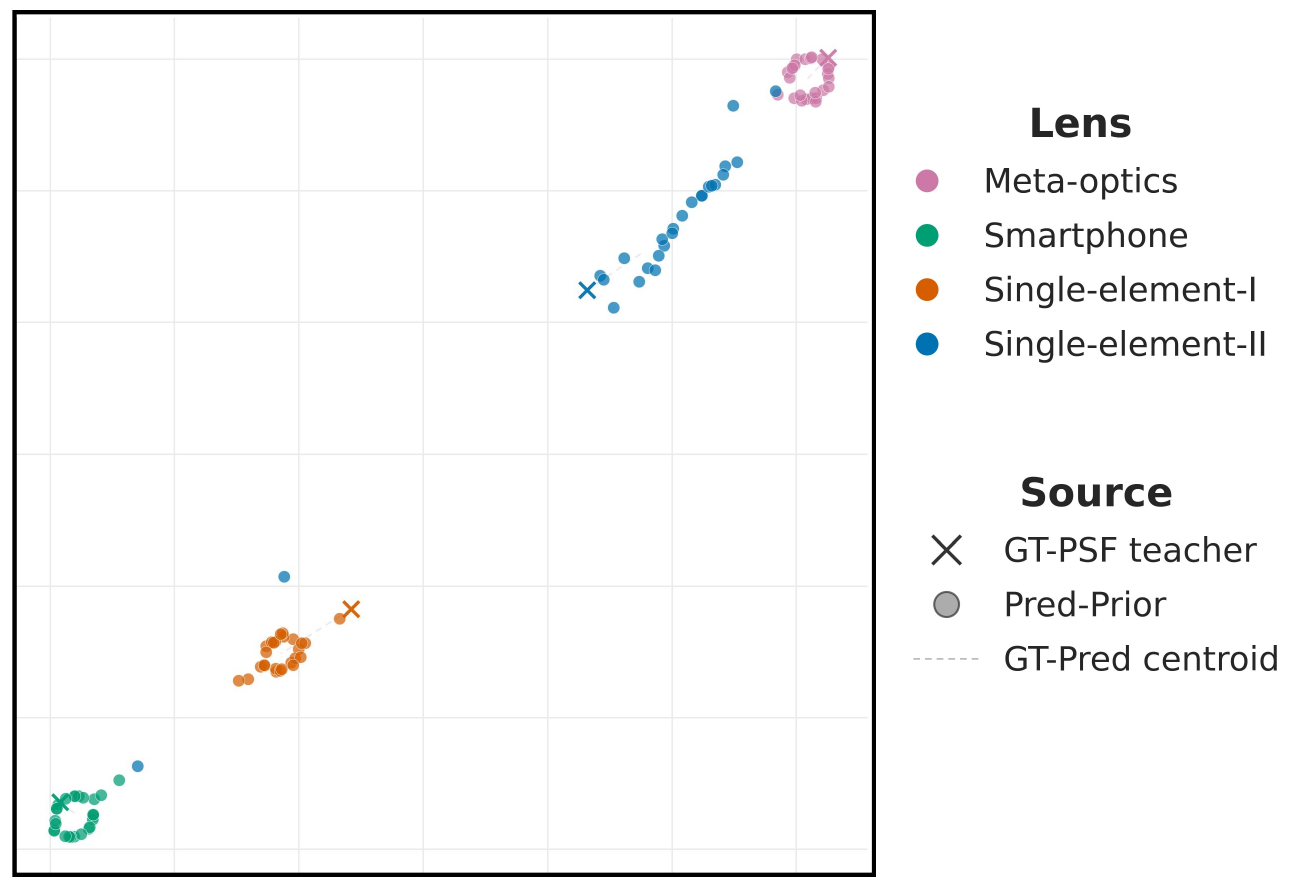}
  % \vspace{-2em}
  \caption{\textbf{t-SNE visualization of blind prior retrieval on representative held-out LensLib lenses.} Crosses denote PSF-derived priors, circles denote image-retrieved priors from aberrated observations, and gray lines connect matched GT--prediction pairs.}   
  % \vskip-3ex
  \label{fig:blind_prior_tsne}
  % \vspace{-2.5em}
\end{figure*}

\noindent\textbf{Blind Prior Retrieval.}
We further analyze whether the image-based degradation predictor can retrieve PSF-derived priors from aberrated observations.
Fig.~\ref{fig:blind_prior_tsne} visualizes four representative lenses from the held-out LensLib split, including two single-element lenses, one smartphone lens, and one meta-optical system.
For each lens, the cross denotes its PSF-derived prior, while the circles denote priors retrieved from different aberrated images under the same lens degradation.
The retrieved priors concentrate around their corresponding PSF-derived anchors and remain separated across different optical systems, indicating that the predictor preserves lens-aware degradation information without requiring PSFs at inference.

\noindent\textbf{Frequency-Domain Interpretation of Veiling Glare.}
For clarity, we omit the patch index $p$ in the following derivation.
From Eq.~\ref{eq:compound_forward}, the additional degradation introduced by veiling glare relative
to the aberration-only observation can be written as
\begin{equation}
I_{\mathrm{de}}-I_{\mathrm{ab}}
=
I_{\mathrm{ab}}\odot (T-1)+I_g .
\end{equation}
Taking the Fourier transform gives
\begin{equation}
\mathcal{F}(I_{\mathrm{de}}-I_{\mathrm{ab}})
=
\mathcal{F}(I_{\mathrm{ab}})*\mathcal{F}(T-1)
+
\mathcal{F}(I_g),
\end{equation}
where $*$ denotes convolution in the frequency domain.
The smooth diffuse component $I_g$ naturally emphasizes low frequencies, whereas the multiplicative term couples the transmission variation with image content and is not strictly confined to the low-frequency region.
Empirically, the introduction of veiling glare produces stronger low-frequency amplitude variations in our optical systems.

In contrast, lens aberration follows
\begin{equation}
I_{\mathrm{ab}}=I_c\otimes K,
\qquad
\mathcal{F}(I_{\mathrm{ab}})
=
\mathcal{F}(I_c)\mathcal{F}(K),
\end{equation}
where the optical transfer function $\mathcal{F}(K)$ affects image
structures across different frequency ranges.
Together with the observed spectral differences between aberration-only and
compound degradations, this motivates the use of target-domain amplitude
statistics in FDC and FPD.

\begin{figure*}[!t]
  \centering
  %\vspace{-1em}
  \includegraphics[width=0.99\textwidth]{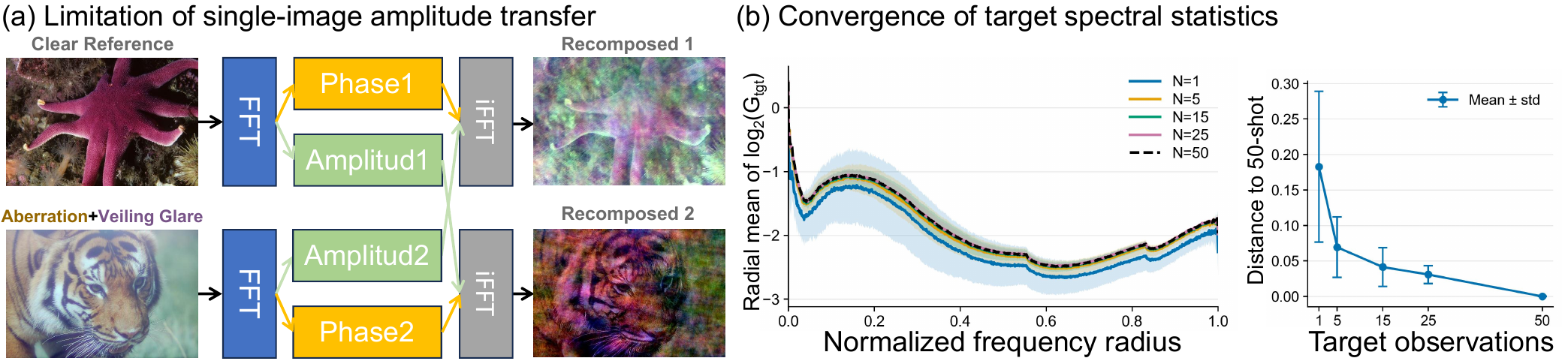}
  % \vspace{-2em}
  \caption{\textbf{Additional analysis of the target spectral statistic.}
  (a) Single-image amplitude transfer introduces content-dependent structures into the recomposed images.
  (b) The estimated target spectral statistic becomes progressively more stable as the number of target observations increases; the right plot reports the distance to the 50-shot statistic.}   
  % \vskip-3ex
  \label{fig:freq_analysis_supp}
  % \vspace{-2.5em}
\end{figure*}

\noindent\textbf{Analysis of Target Spectral Statistics.}
We further analyze why the target statistic is estimated from multiple observations rather than directly transferring the Fourier amplitude of a single target image.
As shown in Fig.~\ref{fig:freq_analysis_supp}(a), single-image amplitude transfer introduces scene-dependent structures into the recomposed images, since the Fourier amplitude contains both degradation and image-content information.
Aggregating statistics over multiple target observations reduces such image-specific variations.
In Fig.~\ref{fig:freq_analysis_supp}(b), the radial profiles progressively converge as more target observations are included, while the distance to the 50-shot statistic consistently decreases.
These results support the use of an aggregated target-system spectral statistic for stable degradation completion.

\noindent\textbf{Effect of Spatial Modulation.}
We further evaluate the spatial modulation mask used in FDC.
As shown in Tab.~\ref{tab:spatial_mask}, introducing spatial variation
maintains comparable reconstruction fidelity while improving perceptual
quality, particularly MUSIQ and Q-Align.
This supports the use of spatial modulation to increase the diversity of the synthesized compound degradation.
\input{Table/spatial_mask_abaltion}

\section{More Results}
\label{sec:More Results}

\subsection{Additional Quantitative Results}
\input{Table/src_enhanced_new_results}
\noindent\textbf{Complete Results under the Target Aberration-Paired Setting.}
We provide the complete quantitative results under the target aberration-paired setting in Tab.~\ref{tab:additional_pair_results}.
The compared methods include cascaded pipelines combining SwinIR~\citep{liang2021swinir} with Flare7K++~\citep{dai2024flare7k++} or DiffDehaze~\citep{wang2025learning}, as well as DiffDehaze, QDMR~\citep{jiang2025qdmr}, DeVeiler~\citep{qian2026deveiler}, and OmniLens~\citep{jiang2026omnilens}.
In addition to the paired \texttt{Screen} domain reported in the main paper, we include all metrics on both the \texttt{Screen} and \texttt{Real} domains.

\input{Table/two_setting_compare_supp}

\input{Table/few_shot}

\input{Table/clear_image_collection_ablation}

\noindent\textbf{Comparison with One-Step Diffusion Restoration Models.}
We further compare \ourframework{} with OSEDiff~\citep{wu2024one} and S3Diff~\citep{zhang2024degradation}, which also employ Stable-Diffusion-based one-step restoration backbones.
Tab.~\ref{tab:supp_sd_comparison} reports results under both the unpaired-target and target aberration-paired settings across the four SL/MRL domains.
\ourframework{} achieves consistently favorable performance, particularly in perceptual and no-reference image quality metrics, providing a comparison with closely related one-step diffusion restoration models.

\noindent\textbf{Target Data Efficiency.}
Tab.~\ref{tab:fewshot_full} provides the complete quantitative results corresponding to the few-shot analysis in the main paper.
Using only five unpaired target observations improves PSNR from $15.075$ to $19.551$ and LPIPS from $0.332$ to $0.277$.
Further increasing the number of target images yields smaller incremental improvements, consistent with the trend observed in the main paper.

\noindent\textbf{Sensitivity to the Clear Reference Dataset.}
The target spectral statistic in Eq.~\ref{eq:target_statistic} is computed using a generic clear-image collection.
We evaluate the sensitivity to this choice on the \texttt{Real}-SL domain by replacing the default DIV2K collection~\citep{timofte2017ntire} with Flickr2K~\citep{lim2017enhanced}.
As shown in Tab.~\ref{tab:clear_dataset_sensitivity}, the results remain comparable between the two clear reference collections, indicating limited sensitivity to the choice of $\mathcal{D}_{\mathrm{clr}}$.

\subsection{Additional Visual Comparisons}

\begin{figure*}[h]
  \centering
  %\vspace{-1em}
  \includegraphics[width=1.0\textwidth]{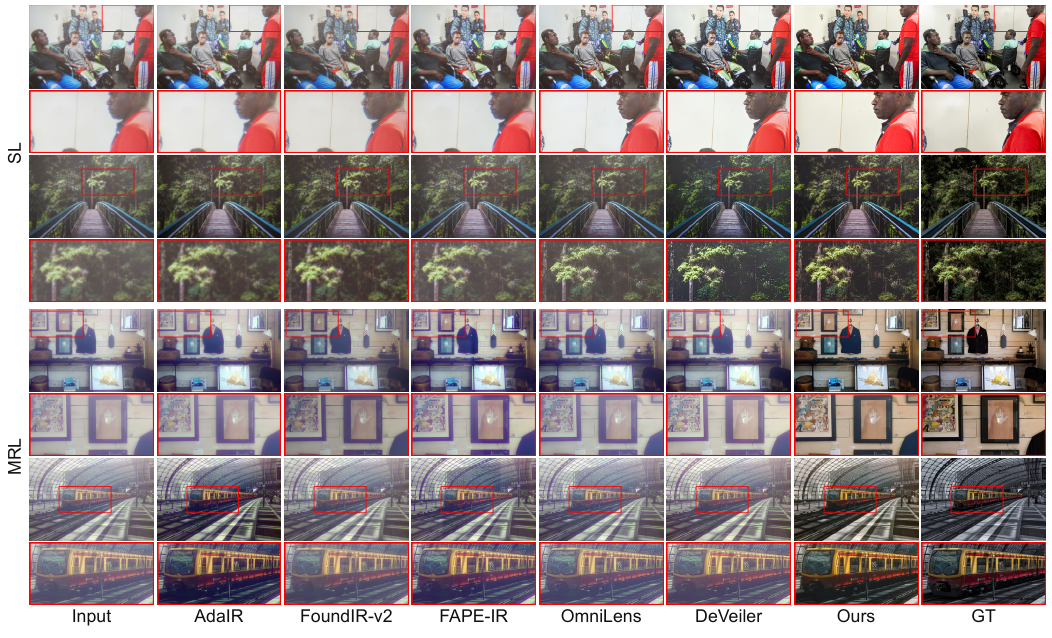}
  % \vspace{-2em}
  \caption{\textbf{Additional visual comparisons on the \texttt{Screen} domain under the unpaired-target setting.} Results are shown for both the SL and MRL optical systems. Zoom in for the best view.}   
  % \vskip-3ex
  \label{fig:unpaired_screen_supp}
  % \vspace{-2.5em}
\end{figure*}

\begin{figure*}[h]
  \centering
  %\vspace{-1em}
  \includegraphics[width=1.0\textwidth]{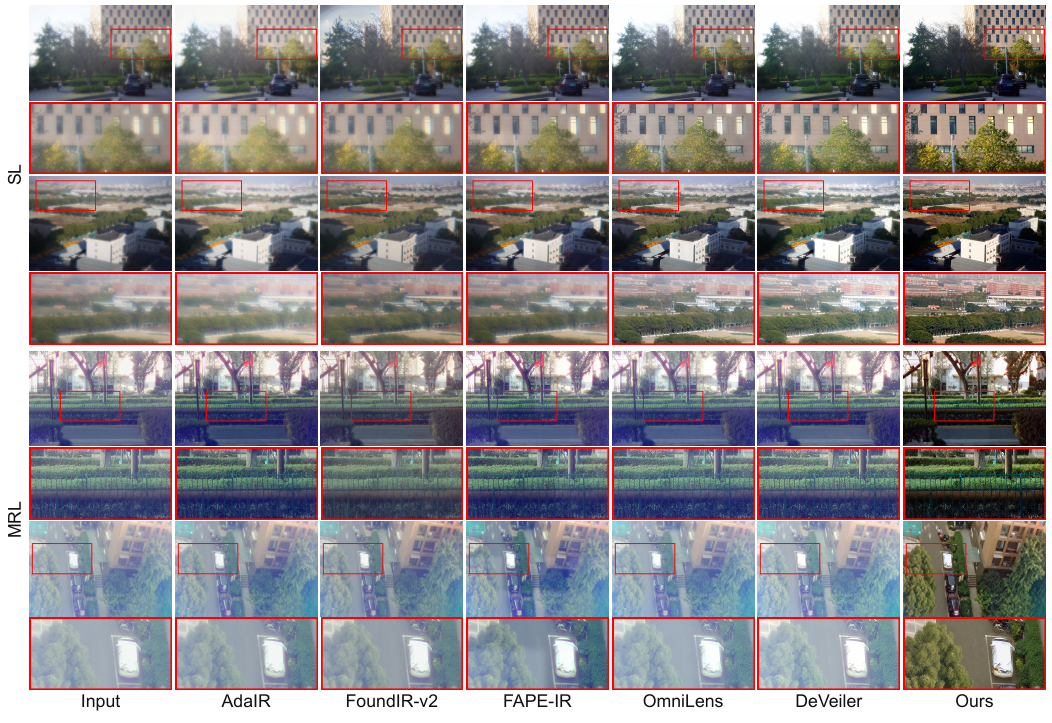}
  % \vspace{-2em}
  \caption{\textbf{Additional visual comparisons on the \texttt{Real} domain under the unpaired-target setting.} Results are shown for both the SL and MRL optical systems. Zoom in for the best view.}   
  % \vskip-3ex
  \label{fig:unpaired_realworld_supp}
  % \vspace{-2.5em}
\end{figure*}

\noindent\textbf{Visual Comparisons under the Unpaired-Target Setting.}
Figs.~\ref{fig:unpaired_screen_supp} and~\ref{fig:unpaired_realworld_supp} provide additional visual comparisons on the \texttt{Screen} and \texttt{Real} domains, respectively.
Across both SL and MRL systems, \ourframework{} better recovers fine structures degraded by aberration while reducing the contrast and color degradation associated with veiling glare.
The improvements remain consistent across controlled screen captures and in-the-wild observations.

\begin{figure*}[h]
  \centering
  %\vspace{-1em}
  \includegraphics[width=1.0\textwidth]{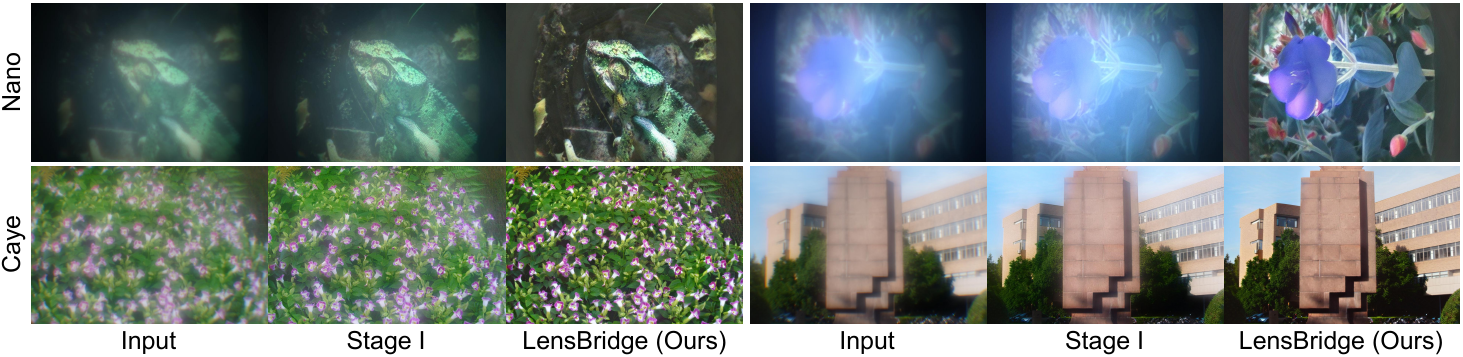}
  % \vspace{-2em}
  \caption{\textbf{Additional visual results on the CAYE lens and nano-optics system.} Stage I denotes the reusable aberration correction foundation before target adaptation, while \ourframework{} denotes the adapted model using unpaired target observations. Zoom in for the best view.}   
  % \vskip-3ex
  \label{fig:nano_caye_results_supp}
  % \vspace{-2.5em}
\end{figure*}

\noindent\textbf{Results on Additional Optical Systems.}
We provide additional results on the CAYE lens and nano-optics system, where the Stage-I foundation is independently adapted to each system using only a few unpaired degraded observations.
As shown in Fig.~\ref{fig:nano_caye_results_supp}, Stage I already alleviates aberration-induced blur and recovers part of the image details, while subsequent target adaptation further improves contrast and color under compound degradation.
The consistent behavior across both systems supports the applicability of the two-stage framework beyond the SL and MRL benchmarks.

\begin{figure*}[h]
  \centering
  %\vspace{-1em}
  \includegraphics[width=1.0\textwidth]{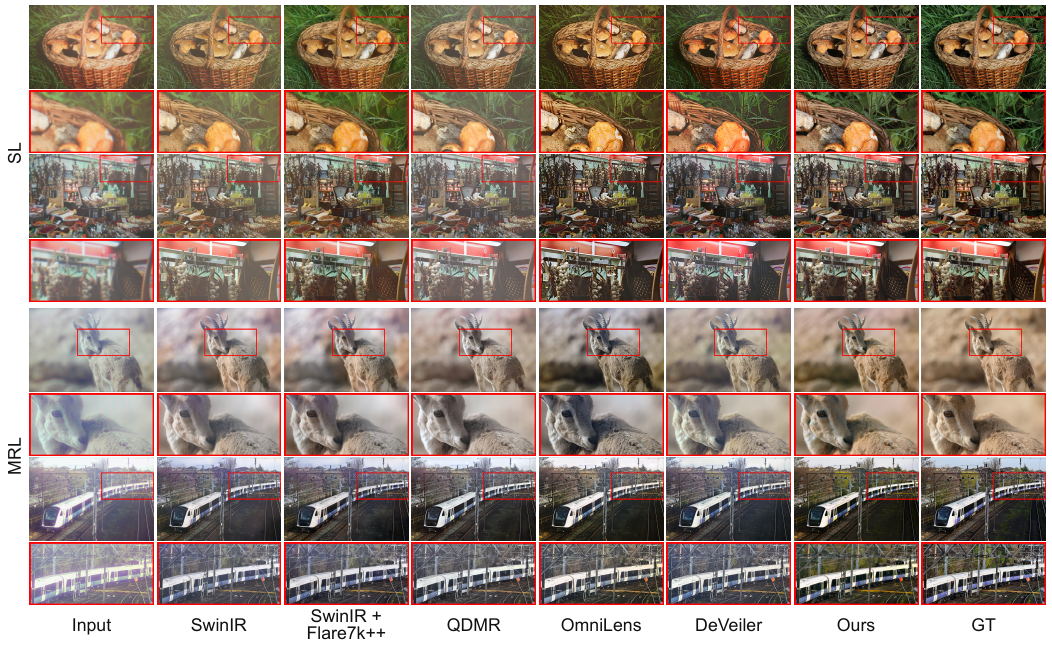}
  % \vspace{-2em}
    \caption{
    \textbf{Additional visual comparisons on the \texttt{Screen} domain under the target aberration-paired setting.}
    Results are shown for both the SL and MRL systems.
    Zoom in for the best view.
    }
  % \vskip-3ex
  \label{fig:paired_target_screen_supp}
  % \vspace{-2.5em}
\end{figure*}

\begin{figure*}[h]
  \centering
  %\vspace{-1em}
  \includegraphics[width=1.0\textwidth]{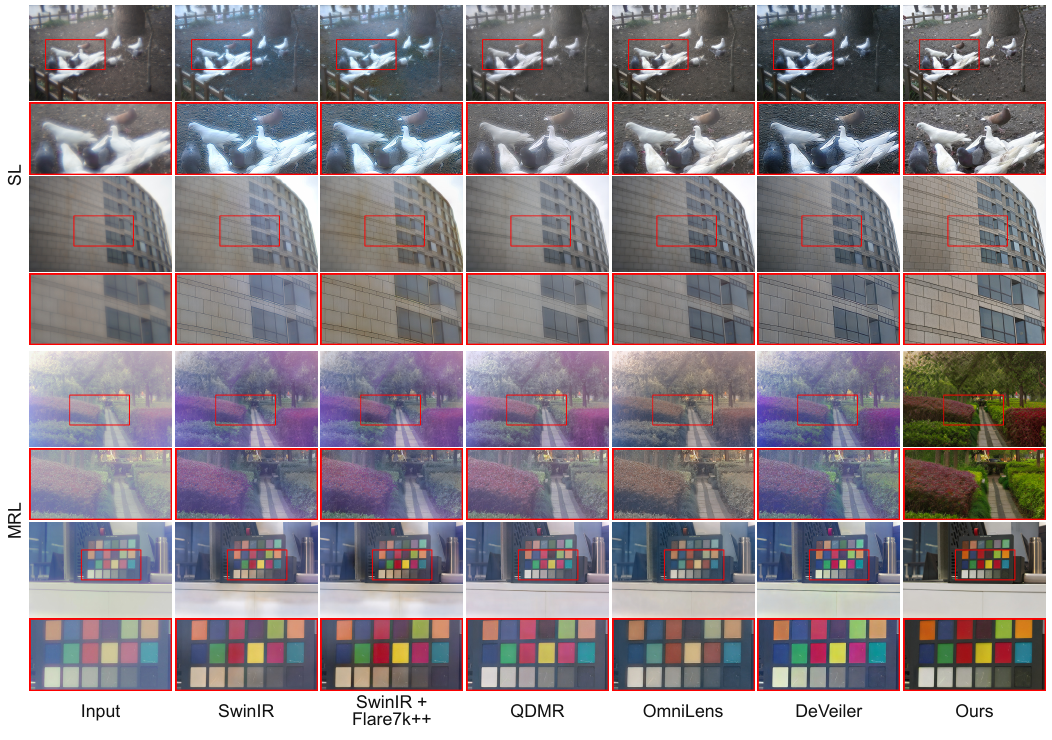}
  % \vspace{-2em}
    \caption{
    \textbf{Additional visual comparisons on the \texttt{Real} domain under the target aberration-paired setting.}
    Results are shown for both the SL and MRL systems.
    Zoom in for the best view.
    }
  % \vskip-3ex
  \label{fig:paired_target_real_supp}
  % \vspace{-2.5em}
\end{figure*}

\noindent\textbf{Visual Comparisons under the Target Aberration-Paired Setting.}
We provide additional visual comparisons under the target aberration-paired setting in Figs.~\ref{fig:paired_target_screen_supp} and~\ref{fig:paired_target_real_supp}.
Across both SL and MRL systems, \ourframework{} produces sharper structures and improved contrast compared with existing restoration and adaptation methods.
These results complement the quantitative comparisons under the same setting, with consistent improvements across different optical systems.

\begin{figure*}[h]
  \centering
  %\vspace{-1em}
  \includegraphics[width=1.0\textwidth]{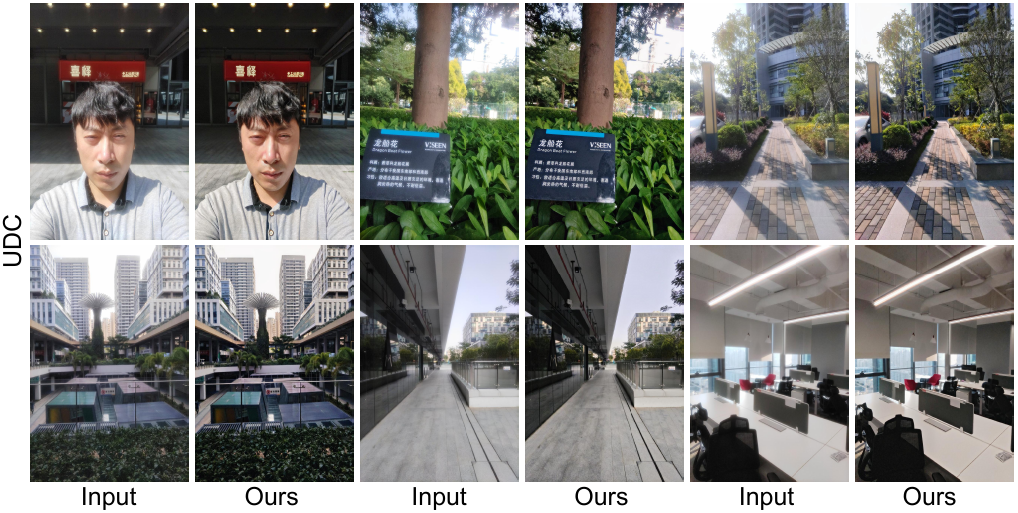}
  % \vspace{-2em}
    \caption{
    \textbf{Visual results on real UDC images} from~\cite{feng2021removing}.
    \ourframework{} provides moderate restoration of contrast and visibility under UDC degradation.
    Zoom in for the best view.
    }
  
  % \vskip-3ex
  \label{fig:udc_results}
  % \vspace{-2.5em}
\end{figure*}

\noindent\textbf{Additional Results on Real UDC Images.}
We further evaluate \ourframework{} on real UDC images from~\cite{feng2021removing}.
UDC images exhibit severe contrast attenuation and haze-like appearance caused by complex optical artifacts, providing an additional out-of-distribution case beyond the current LensLib degradation distribution.
As shown in Fig.~\ref{fig:udc_results}, the method improves global contrast and visibility across diverse indoor and outdoor scenes, while fine diffraction artifacts remain challenging due to the absence of diffraction-aware PSFs in the current LensLib distribution.
This experiment provides an additional qualitative evaluation of the proposed foundation-to-compound adaptation paradigm under unseen optical degradations.

\begin{figure*}[h]
  \centering
  %\vspace{-1em}
  \includegraphics[width=0.85\textwidth]{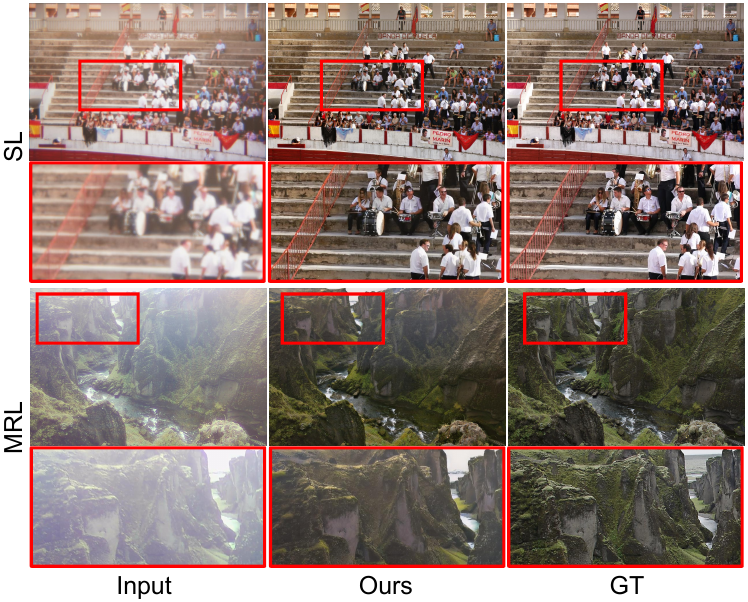}
  % \vspace{-2em}
  \caption{
  \textbf{Failure cases.}
  Fine high-frequency text in SL remains challenging to reconstruct faithfully, while severe overexposure in MRL causes substantial information loss.
  }
  
  % \vskip-3ex
  \label{fig:failure_cases}
  % \vspace{-2.5em}
\end{figure*}

\noindent\textbf{Failure Cases and Future Work.}
Fig.~\ref{fig:failure_cases} shows two representative challenging cases.
For the SL system, fine high-frequency structures such as small text remain difficult to reconstruct accurately, partly due to latent-space compression and the absence of dedicated high-frequency restoration designs.
For the MRL system, severe overexposure produces saturated regions with substantial information loss, preventing faithful recovery of the missing content.
Future work will explore dedicated high-frequency restoration modules, more robust handling of severely saturated observations.

\end{document}

%% file: math_commands.tex
\usepackage{amsmath,amsfonts,bm}

\def\eqref#1{equation~\ref{#1}}
\def\1{\bm{1}}

\DeclareMathAlphabet{\mathsfit}{\encodingdefault}{\sfdefault}{m}{sl}
\SetMathAlphabet{\mathsfit}{bold}{\encodingdefault}{\sfdefault}{bx}{n}

%% file: Tex/Intro.tex
Simplified optical designs enable compact imaging systems for mobile photography~\citep{chen2021extreme_quality,chen_mobile_2023} and emerging meta-optical devices~\citep{li2021meta,chen2024hybrid}, but often leave substantial residual lens aberrations.
Meanwhile, surface imperfections and non-ideal coatings can also induce diffuse stray-light scattering, manifested as \textit{Veiling Glare} (VG)~\citep{qian2026deveiler}.
Lens aberrations introduce spatially varying blur and chromatic dispersion, while veiling glare causes diffuse luminance contamination and contrast reduction.
Their coexistence results in \textit{compound optical degradation}, motivating computational restoration capable of addressing both effects.

\begin{figure*}[!t]
  \centering
  \includegraphics[width=0.99\linewidth]{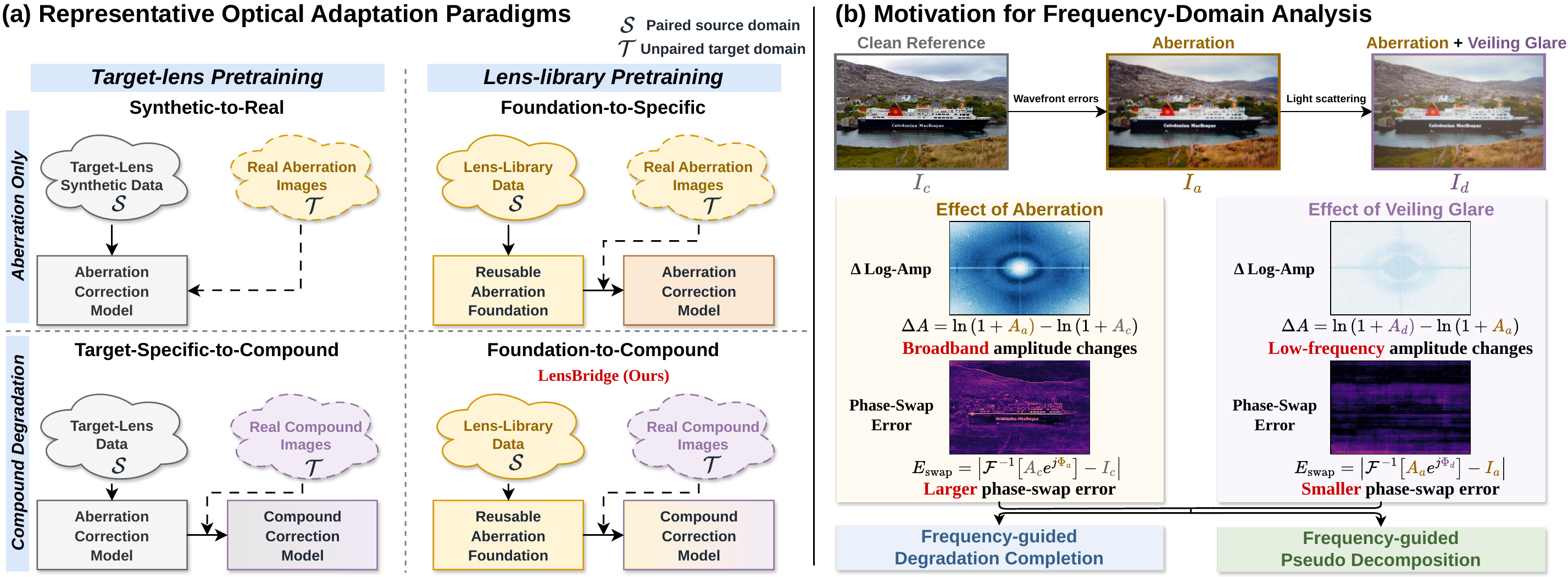}
  \vspace{-1.0em}
  \caption{\textbf{Motivation for \ourframework{}.}
(a) Comparison of optical adaptation paradigms in terms of source pretraining and target degradation. \ourframework{} adapts a reusable aberration correction foundation to target compound degradation using only unpaired target images.
(b) In the illustrated example, \AB{aberration} induces broadband amplitude changes, whereas the additional \VG{veiling glare} produces predominantly low-frequency amplitude changes and a smaller phase-swap reconstruction error in the illustrated example, motivating the frequency-guided designs of FDC and FPD.
} 
  \label{fig:teaser}
  \vspace{-2.0em}
\end{figure*}

Existing computational restoration for optical degradation has primarily focused on lens aberrations.
Conventional learning-based approaches typically rely on target-system synthetic pairs generated from Point Spread Functions (PSFs)~\citep{chen2021extreme_quality,chen2021optical}.
As illustrated by the \textit{synthetic-to-real} paradigm in Fig.~\ref{fig:teaser}(a), such models are trained on target-lens synthetic data and then transferred to real aberrated images~\citep{jiang2025qdmr}.
Acquiring target-lens PSFs and constructing synthetic training pairs for new optical system make deployment costly.
Pretraining on large-scale Lens Libraries (LensLib) alleviates this burden by learning reusable correction models across diverse aberration distributions~\citep{gong2024physics,jiang2026omnilens,qian2026foundcac}.
Among these methods, OmniLens~\citep{jiang2026omnilens} follows the \textit{foundation-to-specific} paradigm in Fig.~\ref{fig:teaser}(a), adapting a LensLib foundation to unseen lenses with unpaired aberrated images and a dark-channel prior.

However, these reusable foundations are learned within the aberration-only degradation space of LensLib and therefore cannot account for veiling glare.
Extending them to compound degradation remains challenging, as physically accurate veiling glare simulation requires optomechanical models and computationally expensive non-sequential ray tracing~\citep{raskar2008glare,hullin2011physically}, making reliable target-system compound pairs difficult to obtain.
A recent solution is target-specific compound degradation synthesis.
DeVeiler~\citep{qian2026deveiler} follows the \textit{target-specific-to-compound} paradigm in Fig.~\ref{fig:teaser}(a), using target-lens aberration pairs and unpaired compound observations to synthesize realistic lens-specific compound degradations for joint restoration.
However, its synthesis remains tailored to each target system rather than reusing an aberration correction foundation across lenses.
In contrast, we pursue a \textit{foundation-to-compound} paradigm, where reusable aberration knowledge is retained from LensLib and only a few unpaired compound observations are required to adapt the foundation to a new optical system.

To enable such adaptation, we examine the two degradation mechanisms in the frequency domain, where PSF-induced blur and diffuse stray-light effects exhibit different spectral responses.
PSF-induced aberration modifies image content across a broad frequency range, whereas veiling glare primarily affects low-frequency amplitude components, as illustrated in Fig.~\ref{fig:teaser}(b).
This distinction motivates using target-domain low-frequency amplitude statistics to characterize the degradation information missing from the aberration-only foundation.

We therefore propose \ourframework{}, a two-stage framework that first establishes a reusable aberration correction foundation and then adapts it to compound optical degradation.
In Stage I, we establish a PSF-aware one-step diffusion foundation for reusable aberration correction.
LensLib PSFs are encoded into a discrete degradation prior space, while image-conditioned retrieval enables the corresponding priors to be inferred directly from aberrated observations.
The retrieved priors condition the diffusion restoration backbone, retaining PSF-aware degradation guidance without requiring explicit PSFs at inference.

In Stage II, we adapt this foundation to target compound degradation using only a few unpaired target observations.
Based on the above frequency distinction, we estimate a shared target-system Fourier-amplitude statistic to guide adaptation at both the data and model levels.
At the data level, Frequency-guided Degradation Completion (FDC) performs data-level adaptation by transferring target-domain low-frequency degradation characteristics to LensLib aberrated images while preserving their aberration-related structures and clean counterparts, thereby constructing target-like compound training pairs without learning an additional degradation generator.
At the model level, Frequency-guided Pseudo Decomposition (FPD) uses the same frequency statistic to construct aberration-dominant and VG-dominant pseudo observations, allowing the two adaptation branches to model their distinct degradation characteristics separately.
Together, FDC provides target-like training supervision, while FPD separately conditions the aberration and VG adaptation branches.
Our main contributions are summarized as follows:
\begin{compactitem}
    \item We propose \ourframework{}, a two-stage framework that first establishes a reusable LensLib-based aberration correction foundation and then adapts it to compound aberration and veiling-glare degradation using only a few unpaired target observations.

    \item We introduce FDC and FPD, which leverage a shared target-domain Fourier-amplitude statistic to construct target-like compound training pairs and aberration-/VG-dominant pseudo observations, respectively.
    
    \item Extensive experiments across multiple optical systems demonstrate the effectiveness of \ourframework{} for joint aberration correction and veiling glare removal.

\end{compactitem}

%% file: Tex/Related_Work.tex
\noindent\textbf{Lens Aberration Correction.}
Computational Aberration Correction (CAC) aims to restore images degraded by residual lens aberrations.
Early approaches relied on deconvolution or PSF-based inverse filtering~\citep{wiener1949extrapolation,fish1995blind,schuler2012blind}.
With the emergence of deep learning, learning-based CAC methods~\citep{chen2021extreme_quality,chen2021optical,jiang2024minimalist,luo2024correcting,zhang2025degradation,mandal2025enabling} learn restoration mappings from simulated optical degradations and achieve improved performance.
However, these methods typically depend on lens-specific simulation pipelines and suffer from limited generalization to unseen optical systems.
Recent works explore reusable CAC models through large-scale lens library pre-training~\citep{jiang2026omnilens,qian2026foundcac}, where diverse aberration distributions improve cross-lens generalization and support adaptation to new lenses.
Nevertheless, existing LensLib-based frameworks remain confined to PSF-induced aberrations and do not address compound degradations involving veiling glare.

\noindent\textbf{Veiling Glare Removal.}
Veiling glare is a diffuse degradation caused by stray-light scattering inside optical systems, leading to luminance contamination and contrast reduction.
Unlike lens flare, which typically appears as localized structures such as bright spots and streaks~\citep{wu2021how_to_train,dai2022flare7k,dai2023nighttime,dai2024flare7k++,vasluianu2024sfnet,zhu2025pbfg}, veiling glare is spatially distributed and dominated by low-frequency variations.
Traditional approaches rely on physical modeling or image decomposition, including Glare Spread Function (GSF) based inverse filtering~\citep{talvala2007veiling} and glare-component decomposition~\citep{zhang2018single}, but often require restrictive assumptions, calibration, or multiple observations.
Recent data-driven methods~\citep{shoshin2021veiling} use simplified flare or scattering models for training-data synthesis, leaving a substantial gap from lens-specific real degradations.
DeVeiler~\citep{qian2026deveiler} addresses this limitation through target-specific compound degradation synthesis, but its degradation modeling must be repeated for each new optical system.
In contrast, \ourframework{} reuses a LensLib-based aberration correction foundation and adapts it to compound degradation using only unpaired target observations.

\noindent\textbf{Frequency-Guided Degradation Adaptation.}
Frequency-domain statistics have been widely explored for domain adaptation and image restoration.
FDA transfers low-frequency Fourier amplitudes between source and target domains for unsupervised semantic segmentation~\citep{yang2020fda}, while FCA exploits frequency consistency to reduce the synthetic-to-real gap in real-world super-resolution~\citep{ji2021frequency}.
Frequency-aware dehazing methods further exploit amplitude-spectrum discrepancies between synthetic and real degradations for adaptation or restoration~\citep{yu2022source,liu2025frequency}.
These methods demonstrate the effectiveness of frequency statistics for characterizing domain shifts, but mainly use them for global source-to-target alignment or adaptation within a single degradation type.
In contrast, \ourframework{} uses an aggregated target-domain Fourier-amplitude statistic to complete the missing target degradation characteristics associated with VG while preserving the existing LensLib aberration structure, and further reuses the same statistic to construct aberration-dominant and VG-dominant pseudo observations.

%% file: Tex/Method.tex
\subsection{Problem Formulation}
\label{sec:problem}

We aim to recover a latent clean image $I_c$ from a compound-degraded image $I_{de}$ captured by a target optical system.
Following~\citep{qian2026deveiler}, we model the joint effects of spatially varying lens aberration and veiling glare.
Approximating the PSF as shift-invariant within each spatial patch~\citep{li2021universal,luo2024correcting,yang2025efficient}, we express the degradation independently for each color channel as:
% \vspace{-1mm}
\begin{equation}
    I_{de}^{p}
    =
    \underbrace{
        \left(I_c^{p}\otimes K^{p}\right)
    }_{I_{ab}^{p}}
    \odot T^{p}
    + I_g^{p},
    \label{eq:compound_forward}
\end{equation}
% \vspace{-1mm}
where $p$ indexes spatial patches, $\otimes$ denotes convolution, and $\odot$ denotes element-wise multiplication.
$K^{p}$ is the local PSF, $I_{ab}^{p}$ is the aberrated patch, $T^{p}$ models contrast attenuation, and $I_g^{p}$ represents additive glare~\citep{zhang2018single,shoshin2021veiling}.
The color-channel index is omitted for clarity.
We use a LensLib source dataset $\mathcal{D}_s=\{(I_{ab,i}^{s},I_{c,i}^{s})\}_{i=1}^{N_s}$, with a corresponding PSF for each pair, and a small set of unpaired compound-degraded target images $\mathcal{D}_t=\{I_{de,j}^{t}\}_{j=1}^{N_t}$.
Here, $s$ and $t$ denote the source and target domains, respectively, and $N_t\ll N_s$.
Our objective is to first establish a reusable aberration correction foundation from $\mathcal{D}_s$ and then adapt it to target compound degradation using only the unpaired observations in $\mathcal{D}_t$.

\begin{figure*}[!t]
  \centering
  %\vspace{-1em}
  \includegraphics[width=0.99\textwidth]{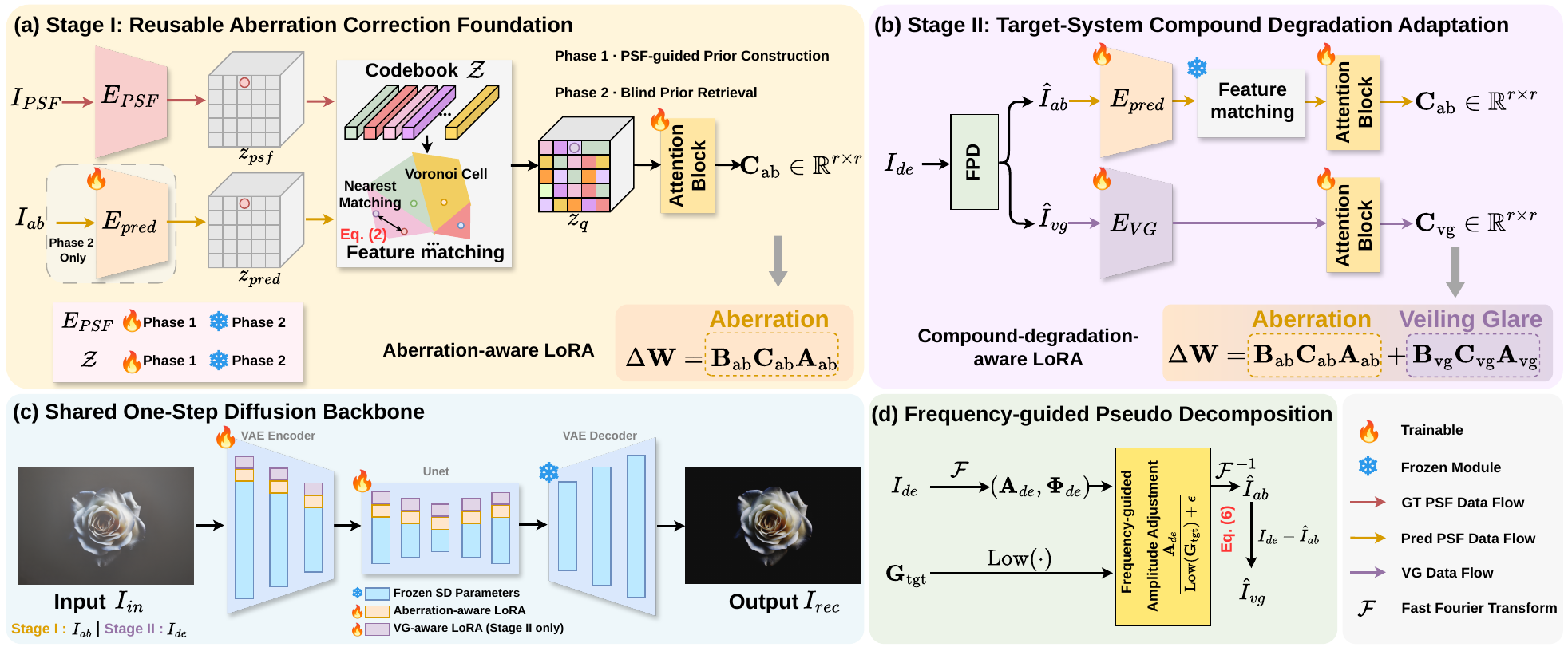}
  \vspace{-1.0em}
    \caption{
    \textbf{Overview of \ourframework{}.}
    \ourframework{} consists of two stages:
    (a) a reusable PSF-aware aberration correction foundation, where discrete degradation priors are constructed from LensLib PSFs and retrieved directly from aberrated observations;
    and (b) target-system compound degradation adaptation, where aberration-dominant and VG-dominant pseudo observations separately condition the two adaptation branches.
    The shared one-step diffusion backbone and the frequency-guided pseudo decomposition mechanism are detailed in (c) and (d), respectively.
    }
  % \vskip-3ex
  \label{fig:LensBridge}
  \vspace{-1.0em}
\end{figure*}

\subsection{Framework Overview}
\label{sec:framework}

As illustrated in Fig.~\ref{fig:LensBridge}, \ourframework{} consists of two stages: establishing a reusable aberration correction foundation and adapting it to target-system compound degradation.
We adopt SD-Turbo~\citep{sauer2024adversarial}, distilled from Stable Diffusion~\citep{rombach2022high}, as the one-step diffusion backbone.
In Stage I, we construct discrete aberration priors from LensLib PSFs and train an image-based predictor to retrieve them from aberrated images.
These priors condition low-rank adaptation (LoRA)~\citep{hu2021lora} of the diffusion backbone.
In Stage II, we estimate a shared target-system Fourier-amplitude statistic from a few unpaired target observations.
Frequency-guided Degradation Completion (FDC) uses this statistic to construct target-like compound training pairs, while Frequency-guided Pseudo Decomposition (FPD) constructs aberration-dominant and VG-dominant pseudo observations to condition separate adaptation branches.

\subsection{Stage I: Reusable Aberration Correction Foundation}
\label{sec:aberration_foundation}

Aberration correction benefits from explicit PSF guidance, whereas target-system PSFs are typically unavailable at inference. We therefore construct a PSF-derived discrete prior space and train the Stage-I foundation in two phases. 
In Phase 1, \textit{PSF-guided Prior Construction}, source-domain PSFs are encoded into discrete aberration priors to condition the diffusion backbone. 
In Phase 2, \textit{Blind Prior Retrieval}, this prior space is fixed and an image-based predictor learns to retrieve the corresponding priors from aberrated observations by matching its predictions to PSF-derived representations during training, enabling PSF-aware correction without requiring PSFs at inference.

\noindent\textbf{Phase 1: PSF-guided Prior Construction.}
Following the discrete PSF representation strategy of FoundCAC~\citep{qian2026foundcac}, we employ vector quantization to construct the PSF-derived prior space.
Specifically, a PSF encoder $E_{\mathrm{PSF}}$ maps a PSF representation
$I_{\mathrm{PSF}}\in\mathbb{R}^{H\times W\times N_p}$ to a continuous representation
$\mathbf{z}_{\mathrm{psf}}\in\mathbb{R}^{H_z\times W_z\times C_z}$, which is quantized using a learnable codebook $\mathcal{Z}$ to obtain
$\mathbf{z}_{\mathrm{psf}}^{q}\in\mathbb{R}^{H_z\times W_z\times C_z}$.
The resulting codebook defines a discrete PSF-derived prior space for PSF-guided conditioning in Phase 1 and image-based retrieval in Phase 2, where image-predicted representations are matched to the corresponding PSF-derived priors during training.
We regularize the encoder and codebook with the vector-quantization objective:
\begin{equation}
    \mathcal{L}_{\mathrm{VQ}}
    =
    \left\|
        \operatorname{sg}[\mathbf{z}_{\mathrm{psf}}]
        - \mathbf{z}_{\mathrm{psf}}^{q}
    \right\|_2^2
    +
    \beta
    \left\|
        \mathbf{z}_{\mathrm{psf}}
        - \operatorname{sg}[\mathbf{z}_{\mathrm{psf}}^{q}]
    \right\|_2^2,
    \label{eq:ab_vq_loss}
\end{equation}
where $\operatorname{sg}[\cdot]$ denotes stop-gradient and $\beta$ controls the commitment term.
The first term updates the selected codebook representations toward the encoder outputs, while the second encourages the encoder outputs to remain close to their assigned codes.

Building on degradation-aware LoRA~\citep{zhang2024degradation}, we condition the low-rank update on PSF-derived discrete priors.
For a pretrained weight
$\mathbf{W}_0\in\mathbb{R}^{d_{\mathrm{out}}\times d_{\mathrm{in}}}$,
standard LoRA uses two low-rank matrices
$\mathbf{B}_{\mathrm{ab}}\in\mathbb{R}^{d_{\mathrm{out}}\times r}$
and
$\mathbf{A}_{\mathrm{ab}}\in\mathbb{R}^{r\times d_{\mathrm{in}}}$,
where $r\ll \min(d_{\mathrm{out}},d_{\mathrm{in}})$.
An attention block maps the quantized prior $\mathbf{z}_{\mathrm{psf}}^{q}$ to a PSF-conditioned modulation matrix
$\mathbf{C}_{\mathrm{ab}}\in\mathbb{R}^{r\times r}$, yielding:
\begin{equation}
    \mathbf{W}
    =
    \mathbf{W}_{0}
    +
    \mathbf{B}_{\mathrm{ab}}
    \mathbf{C}_{\mathrm{ab}}
    \mathbf{A}_{\mathrm{ab}}.
    \label{eq:ab_directlora}
\end{equation}
Compared with the standard LoRA update
$\mathbf{B}_{\mathrm{ab}}\mathbf{A}_{\mathrm{ab}}$,
$\mathbf{C}_{\mathrm{ab}}$ dynamically modulates the low-rank update according to the supplied aberration prior, enabling PSF-conditioned adaptation.
Image reconstruction is supervised by a weighted combination of L1, perceptual, and adversarial losses, denoted as $\mathcal{L}_{\mathrm{rec}}$.
Jointly optimizing $\mathcal{L}_{\mathrm{rec}}$ and $\mathcal{L}_{\mathrm{VQ}}$ encourages the discrete PSF-derived priors to provide effective conditioning for aberration restoration while regularizing their codebook assignments.

\noindent\textbf{Phase 2: Blind Prior Retrieval.}
To enable image-only inference while retaining PSF-derived degradation guidance, we train an RGB degradation predictor $E_{\mathrm{pred}}$ to retrieve the discrete priors established in Phase 1 directly from an aberrated image $I_{ab}$.
To supervise this retrieval, the PSF encoder $E_{\mathrm{PSF}}$ and codebook $\mathcal{Z}$ are kept frozen.
During training, the corresponding PSF representation is encoded and quantized to obtain the teacher prior $\mathbf{z}_{\mathrm{psf}}^{q}$.
A feature-matching loss $\mathcal{L}_{\mathrm{FM}}$~\citep{chen2022real} aligns the predictor's pre-quantization feature $\mathbf{z}_{\mathrm{pred}}$ with $\mathbf{z}_{\mathrm{psf}}^{q}$.
The predicted feature is then quantized by the frozen codebook to obtain $\mathbf{z}_{\mathrm{pred}}^{q}$ for restoration conditioning.
The retrieval stage is optimized with $\mathcal{L}_{\mathrm{rec}}+\mathcal{L}_{\mathrm{FM}}$.
At inference, $E_{\mathrm{pred}}$ retrieves the aberration prior directly from the observed image without requiring the PSF.

\subsection{Stage II: Target-System Compound Degradation Adaptation}
\label{sec:target_adaptation}

In Stage I, we learn a reusable aberration correction foundation for PSF-induced degradation from LensLib.
However, the source distribution does not cover target-specific factors (\textit{i.e.}, veiling glare) beyond the PSF space.
The two degradations arise from different physical mechanisms and therefore exhibit distinct spectral responses.
As illustrated in Fig.~\ref{fig:teaser}(b), PSF-induced aberration modulates a broad frequency range, whereas the spatially smooth contamination of veiling glare mainly affects low-frequency amplitudes.
We therefore estimate a shared target-system Fourier-amplitude statistic from a few unpaired observations to guide data-level degradation completion and model-level pseudo decomposition.

\noindent\textbf{Frequency-guided Degradation Completion.}
Directly transferring the Fourier amplitude of a single target image is unreliable because its spectrum is entangled with scene content.
To obtain a target-system cue less sensitive to image-specific content, we aggregate Fourier-amplitude statistics over a few unpaired target observations.
The resulting statistic is then used to complete the missing target degradation on LensLib aberrated images rather than regenerate the full degradation from clean images.
Given a small target set $\mathcal{D}_{t}$ and an auxiliary clear-image collection $\mathcal{D}_{\mathrm{clr}}$, we first compute their channel-wise mean Fourier amplitudes:
\begin{equation}
    \bar{\mathbf{A}}_{\mathcal{D}}^{c}
    =
    \frac{1}{N_{\mathcal{D}}}
    \sum_{i=1}^{N_{\mathcal{D}}}
    \left|
        \mathcal{F}(I_i^c)
    \right|,
    \qquad
    \mathbf{G}_{\mathrm{tgt}}^{c}
    =
    \frac{
        \bar{\mathbf{A}}_{\mathcal{D}_{t}}^{c}
    }{
        \bar{\mathbf{A}}_{\mathcal{D}_{\mathrm{clr}}}^{c}
        +
        \epsilon
    },
    \label{eq:target_statistic}
\end{equation}
where $\mathcal{F}(\cdot)$ denotes the centered 2D Fourier transform, $c$ indexes the color channel, and $\epsilon$ ensures numerical stability.
Averaging across images reduces image-specific spectral variation, allowing $\mathbf{G}_{\mathrm{tgt}}$ to capture the Fourier-amplitude discrepancy shared by the target observations.
LensLib aberrated images $I_{ab}^{s}$ already contain diverse aberration characteristics that should be retained during target adaptation.
We therefore modulate the Fourier amplitude of $I_{ab}^{s}$ using $\mathbf{G}_{\mathrm{tgt}}$ and retain its original phase.
The low-frequency component of the modulated reconstruction is then combined with the original high-frequency component through wavelet decomposition~\citep{wang2024exploiting}:
\begin{equation}
    \tilde{I}_{de}^{s}
    =
    \mathcal{W}_{H}(I_{ab}^{s})
    +
    \mathcal{W}_{L}
    \left(
        \mathcal{F}^{-1}
        \left[
            \left(
                \mathbf{A}_{ab}^{s}
                \odot
                \mathbf{G}_{\mathrm{tgt}}
            \right)
            e^{j\mathbf{\Phi}_{ab}^{s}}
        \right]
    \right),
    \label{eq:degradation_completion}
\end{equation}
where $\mathbf{A}_{ab}^{s}$ and $\mathbf{\Phi}_{ab}^{s}$ denote the Fourier amplitude and phase of $I_{ab}^{s}$, respectively.
During training, we further apply random spatial masks to modulate the transferred target characteristics, increasing spatial degradation diversity.
Keeping the clean reference $I_{c}^{s}$ unchanged yields the target-like training pair $(\tilde{I}_{de}^{s}, I_{c}^{s})$, completing the missing target degradation while preserving the paired supervision from LensLib.

\noindent\textbf{Frequency-guided Pseudo Decomposition.}
Since aberration and veiling glare exhibit distinct degradation characteristics, a shared LoRA condition may entangle their adaptation signals.
We therefore construct aberration-dominant and VG-dominant pseudo observations to condition separate LoRA branches.
For a compound-degraded image $I_{de}$, we suppress the target-specific low-frequency amplitude shift while retaining its original phase:
\begin{equation}
    \hat{I}_{ab}
    =
    \mathcal{F}^{-1}
    \left[
        \frac{
            \mathbf{A}_{de}
        }{
            \operatorname{Low}(\mathbf{G}_{\mathrm{tgt}})
            +
            \epsilon
        }
        e^{j\mathbf{\Phi}_{de}}
    \right],
    \qquad
    \hat{I}_{vg}
    =
    I_{de}
    -
    \hat{I}_{ab},
    \label{eq:pseudo_decomposition}
\end{equation}
where $\operatorname{Low}(\cdot)$ preserves the target modulation within the low-frequency region and sets the remaining frequencies to one.
The resulting $\hat{I}_{ab}$ and $\hat{I}_{vg}$ serve as aberration-dominant and VG-dominant pseudo observations rather than physically exact degradation components, and are only used to provide factor-specific adaptation cues.
The aberration-dominant observation $\hat{I}_{ab}$ is processed by the Stage-I degradation predictor, and the retrieved discrete aberration prior is mapped by the attention block to $\mathbf{C}_{\mathrm{ab}}$.
In parallel, a lightweight VG encoder $E_{\mathrm{VG}}$ maps $\hat{I}_{vg}$ to the VG condition
$\mathbf{C}_{\mathrm{vg}}\in\mathbb{R}^{r\times r}$.
To preserve these distinct adaptation signals while reusing the Stage-I aberration knowledge, we initialize the aberration branch from Stage I and introduce a separate VG branch:
\begin{equation}
    \mathbf{W}
    =
    \mathbf{W}_{0}
    +
    \mathbf{B}_{\mathrm{ab}}
    \mathbf{C}_{\mathrm{ab}}
    \mathbf{A}_{\mathrm{ab}}
    +
    \mathbf{B}_{\mathrm{vg}}
    \mathbf{C}_{\mathrm{vg}}
    \mathbf{A}_{\mathrm{vg}},
    \label{eq:dual_lora}
\end{equation}
where $(\mathbf{A}_{\mathrm{vg}},\mathbf{B}_{\mathrm{vg}})$ are the learnable low-rank matrices of the VG branch.
During Stage II, FDC-generated compound pairs are jointly used with the original LensLib aberration-only pairs for adaptation.
The pretrained diffusion backbone and discrete codebook remain frozen during Stage II.

%% file: Tex/Exp.tex
\subsection{Experiment Settings}

\noindent\textbf{Datasets.}
We train the aberration correction foundation on LensLib~\citep{Sun_2026_CVPR}, which provides paired aberration data and PSFs.
For target adaptation and evaluation, we use the SL and MRL systems from DeVeiler~\citep{qian2026deveiler}, representing a large-aperture single lens and a metasurface-refractive hybrid lens, respectively.
The \texttt{Screen} domain provides controlled monitor captures with $50$ adaptation images and $42/25$ paired test images for SL/MRL, while the \texttt{Real} domain contains in-the-wild captures with $50$ adaptation images and $51/11$ unpaired test images.
All images are $1280\times1920$.
We further evaluate on the commercial minimalist lens CAYE $50\,\mathrm{mm}$ $f/1.4$ and a nano-optics imaging system~\citep{tseng2021neural}.
Our default \emph{unpaired-target setting} uses only unpaired target images.
We also consider a \emph{target aberration-paired setting} with $170/125$ additional aberration-only pairs for SL/MRL to compare with methods requiring target-specific supervision.
For the generic clear-image collection $\mathcal{D}_{\mathrm{clr}}$ in Eq.~\ref{eq:target_statistic}, we use the DIV2K training set.

\noindent\textbf{Implementation Details.}
We use SD-Turbo~\citep{sauer2024adversarial} as the one-step restoration backbone, with rank-$16$ aberration/VG LoRA branches in the VAE encoder and rank-$32$ branches in the diffusion UNet.
In Stage I, each phase is trained for $30$K iterations at $2\times10^{-5}$; in Stage II, we adapt to each target system for $3$K iterations at $1\times10^{-5}$.
Training uses $512\times512$ random crops with batch size $8$ on an NVIDIA A800 GPU.
Further details are provided in the supplementary material.

\input{Table/Src_free_new_results}
\begin{figure*}[!t]
  \centering
  % \vspace{-0.5em}
  \includegraphics[width=1.0\textwidth]{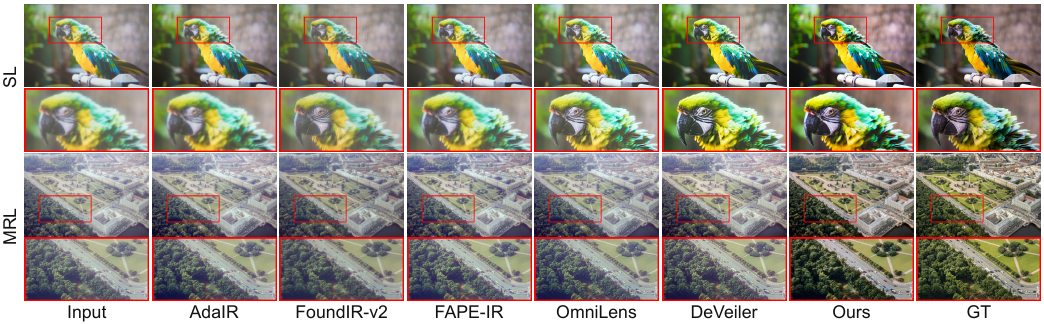}
  \vspace{-2em}
  \caption{\textbf{Visual results on \texttt{Screen} domain.} 
  Zoom in for the best view.}
  % \vskip-3ex
  \label{fig:Visual_results_screen_domain}
  \vspace{-1em}
\end{figure*}

\begin{figure*}[!t]
  \centering
  % \vspace{-0.5em}
  \includegraphics[width=1.0\textwidth]{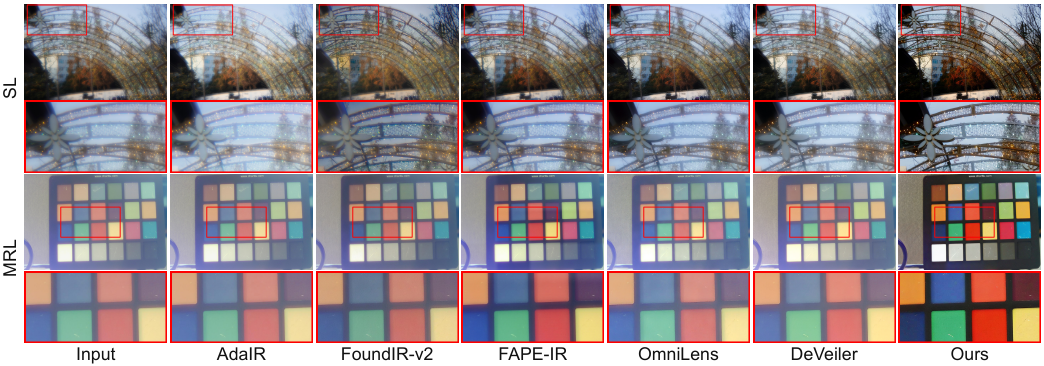}
  \vspace{-2em}
  \caption{\textbf{Visual results on \texttt{Real} domain.} 
  Zoom in for the best view.}
  % \vskip-3ex
  \label{fig:Visual_results_unpaired_realworld}
  %\vspace{-2.0em}
\end{figure*}

\noindent\textbf{Evaluation Metrics.}
We report PSNR, SSIM, and LPIPS~\citep{zhang2018unreasonable} on the paired \texttt{Screen} domain.
For both domains, we report CLIP-IQA~\citep{wang2023exploring}, NIQE~\citep{mittal2012making}, MUSIQ~\citep{ke2021musiq}, and Q-Align~\citep{wu2023q} as no-reference quality metrics.

\subsection{Comparison with State-of-the-Art Methods}
\label{subsec:comparison}
We evaluate \ourframework{} under two target-adaptation settings.
Our primary setting uses only unpaired compound-degraded target images and is also evaluated on additional optical systems.
For fair comparison with DeVeiler~\citep{qian2026deveiler} and its benchmarked methods, we additionally consider the target aberration-paired setting.

\noindent\textbf{Unpaired target setting.}
In the primary setting, target adaptation uses only unpaired compound-degraded images.
We compare against three groups of baselines.
The first is Fast Two-Step~\citep{eboli2022fast}, a blind aberration correction method.
The second comprises four general image restoration methods: DA-CLIP~\citep{luo2023controlling}, AdaIR~\citep{cui2025adair}, FoundIR-v2~\citep{chen2026foundirv2}, and FAPE-IR~\citep{liu2026fape}.
Their pretrained models are directly used for inference.
The third comprises two domain adaptation methods: OmniLens~\citep{jiang2026omnilens} for aberration correction and DeVeiler~\citep{qian2026deveiler} for joint aberration and VG removal.
Both are evaluated under the same unpaired-target setting as \ourframework{}.
As shown in Tab.~\ref{tab:unpaired_target_results}, \ourframework{} performs favorably against existing methods on both \texttt{Screen} and \texttt{Real} domains.
The gains are particularly consistent in LPIPS and no-reference image quality metrics, including over DeVeiler under the same setting.
Visual comparisons in Figs.~\ref{fig:Visual_results_screen_domain} and~\ref{fig:Visual_results_unpaired_realworld} further show that \ourframework{} restores fine details from aberration blur and color contrast degraded by veiling glare on both domains.
These results demonstrate effective adaptation of a reusable aberration foundation to compound degradation with unpaired target observations.

\begin{wrapfigure}{r}{0.48\columnwidth}
    \centering
    \vspace{-3em}
    \includegraphics[width=\linewidth]{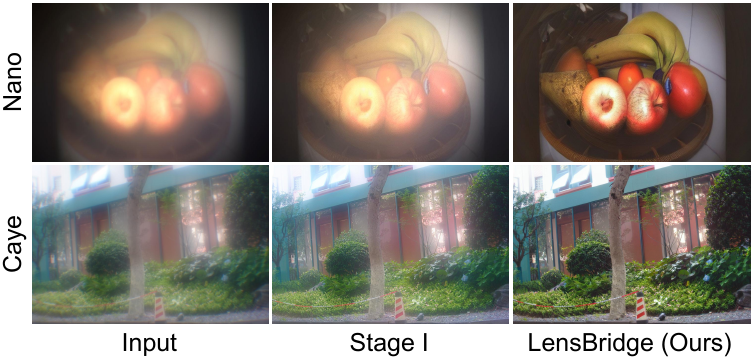}
    \vspace{-2em}
    \caption{\textbf{Results on additional optical systems.}
    \ourframework{} effectively adapts to the CAYE lens and nano-optics system using only a few unpaired target observations.}
    \label{fig:Visual_results_nano_caye}
    \vspace{-1.5em}
\end{wrapfigure}

Beyond SL and MRL, we further evaluate \ourframework{} on the CAYE lens and nano-optics system, independently adapting the foundation to each system with only a few unpaired degraded observations.
As shown in Fig.~\ref{fig:Visual_results_nano_caye}, Stage I already recovers details degraded by aberration, while target adaptation further suppresses veiling glare and restores color and contrast.
These results demonstrate the applicability of the two-stage framework across diverse optical systems.

\input{Table/src_enhanced_new_short_results}

\noindent\textbf{Target aberration-paired setting.}
For comparison with methods benchmarked in DeVeiler~\citep{qian2026deveiler}, we additionally evaluate the target aberration-paired setting, where aberration-clean pairs from the target optical system are available.
We compare with cascaded pipelines combining a retrained SwinIR~\citep{liang2021swinir} with Flare7K++~\citep{dai2024flare7k++} or DiffDehaze~\citep{wang2025learning}, as well as DiffDehaze, QDMR~\citep{jiang2025qdmr}, OmniLens~\citep{jiang2026omnilens}, and DeVeiler~\citep{qian2026deveiler}.
As shown in Tab.~\ref{tab:additional_pair_results_screen}, \ourframework{} achieves the best overall performance across both optical systems and domains.
The cascaded pipelines remain limited, while FDC bridges the target degradation gap and FPD separately conditions the aberration and VG LoRA branches.
Together with the unpaired-target results, these results demonstrate the effectiveness of \ourframework{} under different levels of target-lens supervision.
Additional visual results are provided in the appendix.

\subsection{Ablation Studies}
\label{sec:ablations}

\begin{figure*}[!t]
    \centering

    % ================= Table 3 =================
    \begin{minipage}[t]{0.38\textwidth}
        \centering
        \vspace{0pt}

        \scriptsize
        \setlength{\tabcolsep}{3.2pt}
        \renewcommand{\arraystretch}{0.95}

        \begin{tabular}{l|cccc}
        \toprule
        Method
        & PSNR$\uparrow$
        & LPIPS$\downarrow$
        & MUSIQ$\uparrow$
        & Q-Align$\uparrow$ \\
        \midrule

        CycleGAN
        & 18.85
        & 0.319
        & 56.46
        & 3.998 \\

        HazeGen
        & 20.82
        & 0.269
        & \best{63.26}
        & 4.229 \\

        VeilGen
        & \second{21.56}
        & \second{0.263}
        & \second{62.87}
        & \second{4.244} \\
        
        FDA-style
        & 20.95
        & 0.352
        & 55.35
        & 3.955 \\
        
        \midrule

        \textbf{Ours}
        & \best{23.82}
        & \best{0.253}
        & 61.33
        & \best{4.336} \\

        \bottomrule
        \end{tabular}

        \vspace{-0.5em}

        \captionof{table}{
        Comparison of data synthesis strategies on the \texttt{Screen} SL domain.
        }
        \label{tab:ablation_data_generation}

    \end{minipage}
    \hfill
    % ================= Fig 6 =================
    \begin{minipage}[t]{0.58\textwidth}
        \centering
        \vspace{0pt}

        \includegraphics[width=0.99\linewidth]{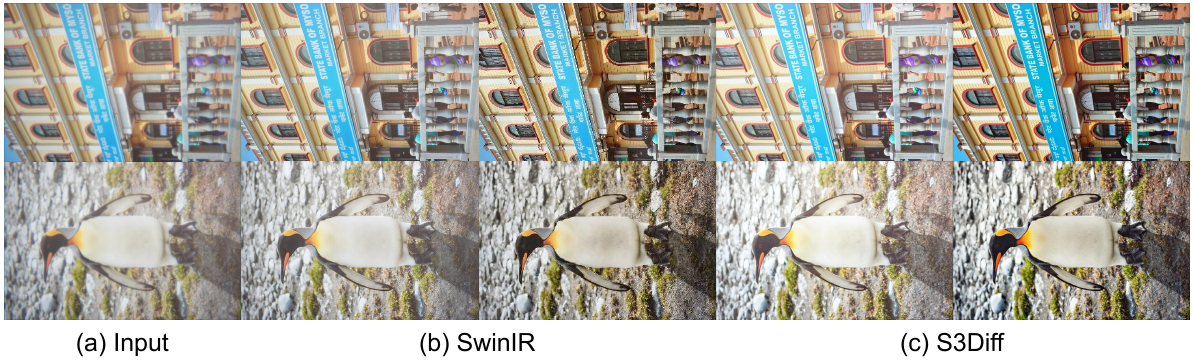}

        \vspace{-1em}

        \captionof{figure}{
        Generality of FDC across restoration backbones on the \texttt{Screen} SL domain.
        }
        \label{fig:Visual_results_data_generalization}

    \end{minipage}

    \vspace{-1.0em}

\end{figure*}

\noindent\textbf{Effectiveness of FDC.}
We adopt the regression restoration model from DeVeiler~\citep{qian2026deveiler} and pretrain it on aberration-only pairs from the \texttt{Screen}-SL dataset.
We compare different synthesis strategies, including CycleGAN~\citep{CycleGAN2017}, HazeGen~\citep{wang2025learning}, VeilGen~\citep{qian2026deveiler}, and FDA~\citep{yang2020fda}, under the same adaptation setting.
As shown in Tab.~\ref{tab:ablation_data_generation}, FDC achieves the best performance by using aggregated target-domain spectral statistics rather than directly transferring a single target observation.
We further evaluate FDC with SwinIR~\citep{liang2021swinir} and S3Diff~\citep{zhang2024degradation}.
As shown in Fig.~\ref{fig:Visual_results_data_generalization}, FDC consistently improves both models, demonstrating its generality across restoration architectures.

\begin{figure*}[!t]
    \centering

    \begin{minipage}[t]{0.51\textwidth}
        \centering
        \vspace{-2pt}

        \scriptsize
        \setlength{\tabcolsep}{1.8pt}
        \renewcommand{\arraystretch}{0.95}

        \begin{tabular}{c|c|cccc|cccc}
            \toprule
            Exp.
            & Steps 
            & FDC
            & AB LoRA
            & VG LoRA
            & FPD
            & PSNR$\uparrow$
            & LPIPS$\downarrow$
            & MUSIQ$\uparrow$
            & Q-Align$\uparrow$ \\
            \midrule

            (a)
            & 30K
            &  &  &  &
            & 15.075
            & 0.332
            & 53.651
            & 3.725 \\

            (b)
            & 30K
            & \checkmark &  &  &
            & 19.722
            & 0.314
            & 54.937
            & 4.055 \\

            (c)
            & 60K
            & \checkmark &  &  &
            & 19.386
            & 0.299
            & 58.057
            & 4.151 \\

            (d)
            & 60K
            & \checkmark
            & \checkmark
            &  &
            & \underline{20.444}
            & \underline{0.268}
            & 64.064
            & 4.400 \\

            (e)
            & 60K
            & \checkmark
            & \checkmark
            & \checkmark
            &
            & 20.316
            & 0.269
            & \underline{65.286}
            & \underline{4.458} \\

            \midrule

            \textbf{(f)}
            & 60K
            & \checkmark
            & \checkmark
            & \checkmark
            & \checkmark
            & \textbf{20.462}
            & \textbf{0.266}
            & \textbf{65.303}
            & \textbf{4.473} \\

            \bottomrule
        \end{tabular}

        \vspace{-0.5em}

        \captionof{table}{
        Component analysis of \ourframework{}.
        }
        \label{tab:component_ablation}

    \end{minipage}
    \hfill
    \begin{minipage}[t]{0.34\textwidth}
        \centering
        \vspace{-2pt}

        \includegraphics[width=0.96\linewidth]{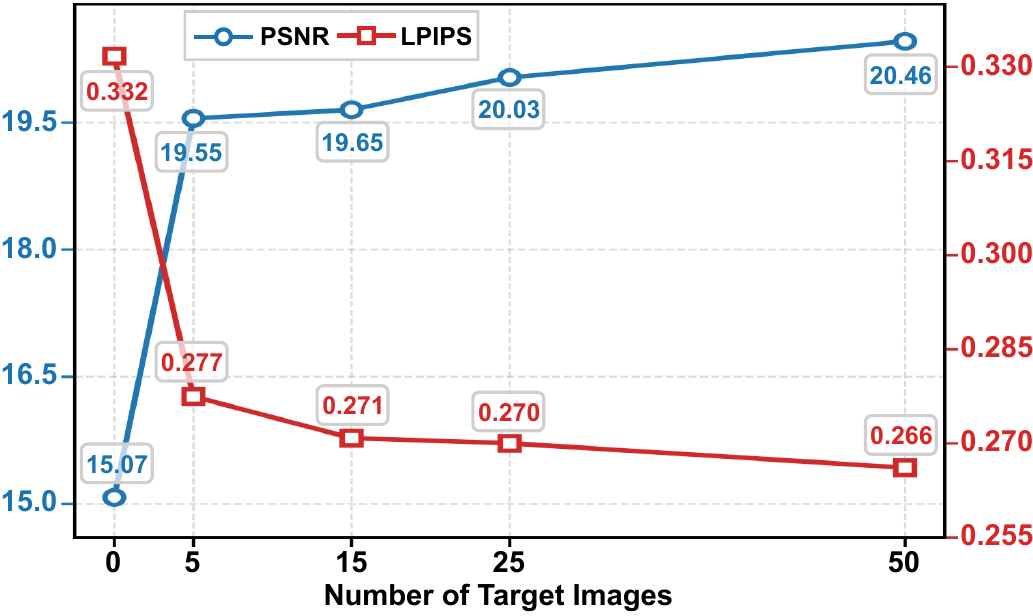}

        \vspace{-1em}

        \captionof{figure}{
        Few-shot adaptation.
        }
        \label{fig:few_shot}

    \end{minipage}

    \vspace{-1.0em}

\end{figure*}

\noindent\textbf{Effectiveness of Individual Components.}
We conduct component ablations on the \texttt{Screen}-SL dataset.
As shown in Tab.~\ref{tab:component_ablation}, FDC-based adaptation in (b) substantially improves the baseline, while longer backbone training in (c) brings limited gains.
The aberration LoRA branch in (d) further improves restoration quality, whereas adding VG LoRA without FPD in (e) provides limited improvement under entangled degradations, suggesting that additional adaptation capacity alone cannot resolve the coupling between different degradation factors.
With FPD, (f) enables separate degradation guidance for the two LoRA branches and achieves the best overall performance.

\noindent\textbf{Few-shot Adaptation.}
We evaluate \ourframework{}'s data efficiency on the \texttt{Screen} SL dataset by varying the number of unpaired target images.
In Fig.~\ref{fig:few_shot}, a few target observations yield substantial gains over the non-adapted model, while additional images yield diminishing improvements.

\begin{figure}[!t]
    \centering
    % \vspace{-0.5em}
    \includegraphics[width=0.99\textwidth]{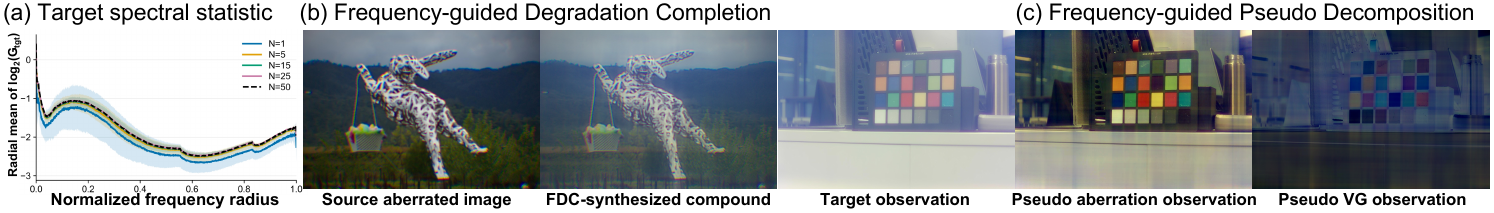}
    \vspace{-1em}
    \caption{
    Analysis of target spectral statistics.
    }
    \vspace{-2.0em}
    \label{fig:frequency_analysis}
\end{figure}
\noindent\textbf{Analysis of Target Spectral Statistics.}
We analyze the $\mathbf{G}_{\mathrm{tgt}}$ on the \texttt{Real} MRL dataset.
In Fig.~\ref{fig:frequency_analysis}(a), aggregating more target observations stabilizes the spectral statistic.
Using this statistic, FDC in Fig.~\ref{fig:frequency_analysis}(b) completes target degradation characteristics while preserving aberration patterns, while FPD in Fig.~\ref{fig:frequency_analysis}(c) constructs aberration- and VG-dominant pseudo observations for separate LoRA branches.
These results support $\mathbf{G}_{\mathrm{tgt}}$ as a target cue for compound degradation adaptation.

%% file: Table/src_free_new_results.tex
\begin{table*}[!t]
\footnotesize
\centering
\setlength{\tabcolsep}{1.05mm}
\renewcommand{\arraystretch}{0.95}
\caption{
\textbf{Quantitative comparisons on the SL and MRL optical systems under the unpaired target setting.}
Methods requiring adaptation use only unpaired target observations.
\red{\textbf{Red}} and \blue{\underline{blue}} indicate the best and the second best performance, respectively. $^{*}$ denotes the pretrained model.
}
\label{tab:unpaired_target_results}
\vskip-2.5ex

\begin{adjustbox}{width=0.99\textwidth}
\begin{tabular}{c c|c|cccc|cc|c}
\toprule

Dataset
&
Metric
&
\methodhead{Fast\\Two-Step$^{*}$}
&
DA-CLIP$^{*}$
&
AdaIR$^{*}$
&
FoundIR-v2$^{*}$
&
FAPE-IR$^{*}$
&
OmniLens
&
DeVeiler
&
\methodhead{LensBridge\\(Ours)}
\\

\midrule

\multirow{7}{*}{\shortstack{\texttt{Screen}\\SL}}
& PSNR $\uparrow$
& 16.62
& 16.74
& 17.24
& 17.40
& 17.72
& 17.67
& \second{19.18}
& \best{20.46}
\\

& SSIM $\uparrow$
& 0.539
& 0.620
& 0.597
& 0.539
& 0.626
& 0.541
& \best{0.678}
& \second{0.652}
\\

& LPIPS $\downarrow$
& 0.676
& 0.428
& 0.582
& 0.465
& 0.487
& 0.430
& \second{0.327}
& \best{0.266}
\\

& CLIP-IQA $\uparrow$
& 0.298
& 0.389
& 0.246
& 0.476
& 0.253
& \second{0.609}
& 0.526
& \best{0.620}
\\

& NIQE $\downarrow$
& 6.908
& 4.630
& 8.661
& 5.373
& 6.508
& \second{4.338}
& 4.856
& \best{3.417}
\\

& MUSIQ $\uparrow$
& 21.86
& 48.48
& 22.54
& 52.45
& 32.73
& 55.66
& \second{57.96}
& \best{65.30}
\\

& Q-Align $\uparrow$
& 2.567
& 3.122
& 2.366
& 3.272
& 2.932
& 3.859
& \second{4.154}
& \best{4.473}
\\

\midrule

\multirow{7}{*}{\shortstack{\texttt{Screen}\\MRL}}
& PSNR $\uparrow$
& 14.94
& 14.94
& \second{18.43}
& 17.74
& 17.23
& 16.07
& 15.46
& \best{19.72}
\\

& SSIM $\uparrow$
& 0.610
& 0.641
& \best{0.705}
& 0.634
& \second{0.681}
& 0.666
& 0.663
& 0.652
\\

& LPIPS $\downarrow$
& 0.503
& 0.484
& 0.466
& 0.479
& 0.472
& 0.453
& \second{0.449}
& \best{0.333}
\\

& CLIP-IQA $\uparrow$
& 0.404
& 0.309
& 0.284
& 0.351
& 0.321
& 0.361
& \second{0.456}
& \best{0.556}
\\

& NIQE $\downarrow$
& \second{4.310}
& 4.360
& 4.582
& 5.062
& 5.208
& 4.376
& 4.908
& \best{3.349}
\\

& MUSIQ $\uparrow$
& 44.49
& 42.64
& 42.22
& 43.23
& 42.15
& 45.75
& \second{48.22}
& \best{60.55}
\\

& Q-Align $\uparrow$
& 3.384
& 3.258
& 3.494
& 3.423
& 3.423
& 3.394
& \second{3.582}
& \best{4.400}
\\

\midrule

\multirow{4}{*}{\shortstack{\texttt{Real}\\SL}}
& CLIP-IQA $\uparrow$
& 0.356
& 0.370
& 0.274
& 0.370
& 0.310
& 0.464
& \second{0.466}
& \best{0.572}
\\

& NIQE $\downarrow$
& 5.817
& 5.465
& 7.146
& 6.231
& 6.155
& \second{4.794}
& 5.201
& \best{4.150}
\\

& MUSIQ $\uparrow$
& 33.58
& 38.07
& 25.86
& 40.14
& 36.57
& 51.73
& \second{54.33}
& \best{60.32}
\\

& Q-Align $\uparrow$
& 2.942
& 3.261
& 2.625
& 3.103
& 3.321
& 3.946
& \second{4.290}
& \best{4.354}
\\

\midrule

\multirow{4}{*}{\shortstack{\texttt{Real}\\MRL}}
& CLIP-IQA $\uparrow$
& 0.388
& 0.297
& 0.342
& 0.347
& 0.314
& 0.350
& \second{0.422}
& \best{0.536}
\\

& NIQE $\downarrow$
& 5.332
& 6.658
& 5.802
& 6.169
& 5.947
& \second{5.299}
& 5.629
& \best{4.363}
\\

& MUSIQ $\uparrow$
& 33.35
& 38.44
& 34.27
& 36.28
& 36.52
& 38.91
& \second{43.79}
& \best{56.30}
\\

& Q-Align $\uparrow$
& 2.687
& 2.612
& 2.762
& 2.800
& \second{2.932}
& 2.720
& 2.846
& \best{3.919}
\\

\bottomrule
\end{tabular}
\end{adjustbox}

\vspace{-3mm}
\end{table*}

%% file: Table/src_enhanced_new_short_results.tex
\begin{table*}[!t]
\footnotesize
\vspace{-5mm}
\centering
\setlength{\tabcolsep}{1.2mm}
\renewcommand{\arraystretch}{0.95}

\caption{
\textbf{Quantitative comparisons on the SL and MRL optical systems in the \texttt{Screen} domain under the target aberration-paired setting.}
\red{\textbf{Red}} and \blue{\underline{blue}} indicate the best and the second-best performance, respectively.
$^{*}$ denotes using the pretrained model.
}
\label{tab:additional_pair_results_screen}
\vskip-2.5ex

\begin{adjustbox}{width=0.99\textwidth}
\begin{tabular}{@{}c c|c|cc|cc|cc|c@{}}
\toprule

Dataset
&
Metric
&
SwinIR
&
\methodhead{SwinIR+\\Flare7K++$^{*}$}
&
\methodhead{SwinIR+\\DiffDehaze$^{*}$}
&
DiffDehaze
&
QDMR
&
OmniLens
&
DeVeiler
&
\methodhead{LensBridge\\(Ours)}
\\

\midrule

\multirow{4}{*}{\shortstack{\texttt{Screen}\\SL}}
& PSNR $\uparrow$
& 18.18
& 21.67
& 19.31
& 18.87
& 18.45
& 20.53
& \second{22.38}
& \best{22.69}
\\

& LPIPS $\downarrow$
& 0.298
& 0.297
& 0.347
& 0.435
& 0.291
& 0.262
& \second{0.261}
& \best{0.202}
\\

& MUSIQ $\uparrow$
& 62.08
& 59.60
& 64.29
& 39.35
& 61.52
& \second{65.21}
& 62.29
& \best{68.20}
\\

& Q-Align $\uparrow$
& 4.091
& \second{4.360}
& 4.318
& 3.843
& 4.087
& 4.221
& 4.299
& \best{4.709}
\\

\midrule

\multirow{4}{*}{\shortstack{\texttt{Screen}\\MRL}}
& PSNR $\uparrow$
& 19.34
& 20.74
& 18.28
& 18.55
& 20.67
& 20.64
& \best{21.57}
& \second{21.46}
\\

& LPIPS $\downarrow$
& 0.354
& 0.336
& 0.392
& 0.426
& 0.315
& \second{0.300}
& 0.301
& \best{0.266}
\\

& MUSIQ $\uparrow$
& 58.72
& 58.22
& \second{64.70}
& 54.96
& 61.29
& 62.84
& 58.43
& \best{67.74}
\\

& Q-Align $\uparrow$
& 4.206
& 4.356
& 4.375
& 4.343
& 4.344
& \second{4.435}
& 4.310
& \best{4.617}
\\

\bottomrule
\end{tabular}
\end{adjustbox}

\vspace{-4mm}
\end{table*}

%% file: Tex/Conclusion.tex
In this paper, we address compound optical degradation involving co-existing aberration and veiling glare in simplified optical systems.
We propose \ourframework{}, which adapts a reusable aberration correction foundation to target-specific compound degradation using only a few unpaired target observations, with a shared Fourier-amplitude statistic guiding both FDC and FPD.
Experimental results across multiple optical systems demonstrate favorable restoration performance, effective few-shot adaptation, and applicability across different restoration backbones.
We hope this work can facilitate efficient adaptation of reusable restoration foundations to diverse optical systems.

\noindent\textbf{Limitations and Future Work.}
\ourframework{} relies on the coverage of the LensLib foundation, and broader target-specific aberration priors could further improve unseen lens adaptation.
The current VG adaptation uses frequency-guided degradation completion rather than a fully calibrated physical model, motivating future exploration of broader optical degradation foundations and physically grounded VG modeling under complex illumination.

%% file: Table/few_shot_numbers.tex
\begin{table}[h]
\footnotesize
\centering
\setlength{\tabcolsep}{3mm}
\renewcommand{\arraystretch}{1.1}

\caption{
Number of unpaired target observations used for few-shot Stage-II adaptation across different optical systems.
}
\label{tab:adaptation_data}

\begin{tabular}{c c c c}
\toprule
Optical System & Domain & \# Target Observations & Purpose \\
\midrule

SL 
& \texttt{Screen}
& 50
& Main benchmark
\\

SL 
& \texttt{Real}
& 50
& Main benchmark
\\

MRL 
& \texttt{Screen}
& 50
& Main benchmark
\\

MRL 
& \texttt{Real}
& 50
& Main benchmark
\\

CAYE
& \texttt{Real}
& 26
& Additional optical system
\\

Nano
& \texttt{Real}
& 3
& Additional optical system
\\

UDC
& \texttt{Real}
& 30
& OOD degradation evaluation
\\

\bottomrule
\end{tabular}
\end{table}

%% file: Table/spatial_mask_abaltion.tex
\begin{table}[t]
\centering
\caption{
Effect of spatial modulation during FDC on \texttt{Screen-SL}.
}
\label{tab:spatial_mask}
\footnotesize
\setlength{\tabcolsep}{5.5pt}
\renewcommand{\arraystretch}{1.05}
\begin{tabular}{ccccc}
\toprule
Spatial Modulation
& PSNR $\uparrow$
& LPIPS $\downarrow$
& MUSIQ $\uparrow$
& Q-Align $\uparrow$ \\
\midrule
$\times$
& \textbf{20.505}
& 0.268
& 63.556
& 4.447 \\
$\checkmark$
& 20.462
& \textbf{0.266}
& \textbf{65.303}
& \textbf{4.473} \\
\bottomrule
\end{tabular}
\end{table}

%% file: Table/src_enhanced_new_results.tex
\begin{table*}[!t]
% \scriptsize
\footnotesize
% \vspace{-5mm}
\centering
\setlength{\tabcolsep}{0.68mm}
\renewcommand{\arraystretch}{1.05}

\caption{
\textbf{Quantitative comparisons on the SL and MRL optical systems with additional target-system aberration-only paired supervision.}
This setting supplements the primary setting with target-system aberration pairs.
\red{\textbf{Red}} and \blue{\underline{blue}} indicate the best and the second-best performance, respectively.
}
\label{tab:additional_pair_results}
% \vskip-2ex

\begin{adjustbox}{width=\textwidth}
\begin{tabular}{@{}c c|c|cc|ccc|c|c@{}}
\toprule

Dataset
&
Metric
&
SwinIR
&
\methodhead{SwinIR+\\Flare7K++}
&
\methodhead{SwinIR+\\DiffDehaze}
&
DiffDehaze
&
QDMR
&
DeVeiler
&
OmniLens
&
\methodhead{LensBridge\\(Ours)}
\\

\midrule

\multirow{7}{*}{\shortstack{\texttt{Screen}\\SL}}
& PSNR $\uparrow$
& 18.18
& 21.67
& 19.31
& 18.87
& 18.45
& \second{22.38}
& 20.53
& \best{22.69}
\\

& SSIM $\uparrow$
& 0.686
& \second{0.723}
& 0.642
& 0.657
& 0.681
& \best{0.729}
& 0.642
& 0.679
\\

& LPIPS $\downarrow$
& 0.298
& 0.297
& 0.347
& 0.435
& 0.291
& \second{0.261}
& 0.262
& \best{0.202}
\\

& CLIP-IQA $\uparrow$
& 0.449
& 0.457
& 0.575
& 0.344
& 0.450
& 0.579
& \second{0.617}
& \best{0.636}
\\

& NIQE $\downarrow$
& 5.400
& 3.996
& 5.873
& 7.843
& 4.262
& 4.163
& \best{3.393}
& \second{3.428}
\\

& MUSIQ $\uparrow$
& 62.08
& 59.60
& 64.29
& 39.35
& 61.52
& 62.29
& \second{65.21}
& \best{68.20}
\\

& Q-Align $\uparrow$
& 4.091
& \second{4.360}
& 4.318
& 3.843
& 4.087
& 4.299
& 4.221
& \best{4.709}
\\

\midrule

\multirow{7}{*}{\shortstack{\texttt{Screen}\\MRL}}
& PSNR $\uparrow$
& 19.34
& 20.74
& 18.28
& 18.55
& 20.67
& \best{21.57}
& 20.64
& \second{21.46}
\\

& SSIM $\uparrow$
& 0.722
& \second{0.745}
& 0.603
& 0.645
& 0.725
& \best{0.746}
& 0.738
& 0.674
\\

& LPIPS $\downarrow$
& 0.354
& 0.336
& 0.392
& 0.426
& 0.315
& 0.301
& \second{0.300}
& \best{0.266}
\\

& CLIP-IQA $\uparrow$
& 0.458
& 0.469
& 0.550
& 0.460
& 0.527
& 0.526
& \second{0.569}
& \best{0.669}
\\

& NIQE $\downarrow$
& 5.495
& 5.225
& 3.831
& 5.157
& 3.761
& 4.510
& \second{3.728}
& \best{3.263}
\\

& MUSIQ $\uparrow$
& 58.72
& 58.22
& \second{64.70}
& 54.96
& 61.29
& 58.43
& 62.84
& \best{67.74}
\\

& Q-Align $\uparrow$
& 4.206
& 4.356
& 4.375
& 4.343
& 4.344
& 4.310
& \second{4.435}
& \best{4.617}
\\

\midrule

\multirow{4}{*}{\shortstack{\texttt{Real}\\SL}}
& CLIP-IQA $\uparrow$
& 0.424
& 0.459
& 0.573
& 0.406
& 0.405
& \best{0.607}
& 0.493
& \second{0.595}
\\

& NIQE $\downarrow$
& 5.710
& 5.616
& 6.122
& 6.467
& 4.770
& \second{4.446}
& 4.692
& \best{4.013}
\\

& MUSIQ $\uparrow$
& 58.18
& 58.21
& \second{60.59}
& 47.50
& 47.64
& 59.78
& 57.27
& \best{63.89}
\\

& Q-Align $\uparrow$
& 3.518
& 3.528
& 3.679
& 3.982
& 3.864
& 3.987
& \second{4.090}
& \best{4.451}
\\

\midrule

\multirow{4}{*}{\shortstack{\texttt{Real}\\MRL}}
& CLIP-IQA $\uparrow$
& 0.374
& 0.387
& 0.437
& 0.428
& 0.376
& \second{0.440}
& 0.421
& \best{0.633}
\\

& NIQE $\downarrow$
& 6.696
& 6.345
& \second{5.226}
& 6.840
& 5.478
& 5.286
& 5.505
& \best{4.341}
\\

& MUSIQ $\uparrow$
& 46.18
& 46.11
& \second{49.19}
& 45.96
& 44.58
& 48.57
& 47.39
& \best{62.87}
\\

& Q-Align $\uparrow$
& 3.191
& 3.319
& 3.542
& 3.476
& 3.337
& \second{3.586}
& 3.338
& \best{4.045}
\\

\bottomrule
\end{tabular}
\end{adjustbox}

\end{table*}

%% file: Table/two_setting_compare_supp.tex
\begin{table*}[!t]
\footnotesize
\centering
\setlength{\tabcolsep}{2.6mm}
\renewcommand{\arraystretch}{0.93}

\caption{
\textbf{Comparison with SD-based one-step diffusion restoration models under two target-adaptation settings.}
Results are reported under the unpaired-target setting and the target aberration-paired setting.
The best and second-best results within each setting are highlighted in
\red{\textbf{Red}} and \blue{\underline{blue}} indicate the best and the second-best performance, respectively.
}
\label{tab:supp_sd_comparison}

\begin{adjustbox}{width=0.98\textwidth}
\begin{tabular}{c c|ccc|ccc}
\toprule

\multirow{2}{*}{Dataset}
&
\multirow{2}{*}{Metric}
&
\multicolumn{3}{c|}{Unpaired-Target}
&
\multicolumn{3}{c}{Target Aberration-Paired}
\\

\cmidrule(lr){3-5}
\cmidrule(lr){6-8}

&
&
OSEDiff
&
S3Diff
&
\methodhead{LensBridge\\(Ours)}
&
OSEDiff
&
S3Diff
&
\methodhead{LensBridge\\(Ours)}
\\

\midrule

\multirow{7}{*}{\shortstack{\texttt{Screen}\\SL}}
& PSNR $\uparrow$
& \best{20.94}
& 20.07
& \second{20.46}
& \second{22.39}
& 21.72
& \best{22.69}
\\

& SSIM $\uparrow$
& \best{0.677}
& \second{0.658}
& 0.652
& \best{0.679}
& 0.625
& \best{0.679}
\\

& LPIPS $\downarrow$
& 0.320
& \second{0.282}
& \best{0.266}
& 0.260
& \second{0.248}
& \best{0.202}
\\

& CLIP-IQA $\uparrow$
& 0.524
& \second{0.593}
& \best{0.620}
& 0.581
& \best{0.659}
& \second{0.636}
\\

& NIQE $\downarrow$
& 4.285
& \second{3.457}
& \best{3.417}
& 3.984
& \second{3.444}
& \best{3.428}
\\

& MUSIQ $\uparrow$
& 60.68
& \best{65.85}
& \second{65.30}
& 66.05
& \second{68.13}
& \best{68.20}
\\

& Q-Align $\uparrow$
& 4.289
& \second{4.440}
& \best{4.473}
& 4.554
& \best{4.749}
& \second{4.709}
\\

\midrule

\multirow{7}{*}{\shortstack{\texttt{Screen}\\MRL}}
& PSNR $\uparrow$
& 18.95
& \second{19.30}
& \best{19.72}
& 20.42
& \second{20.99}
& \best{21.46}
\\

& SSIM $\uparrow$
& \best{0.665}
& \second{0.659}
& 0.652
& \best{0.681}
& 0.664
& \second{0.674}
\\

& LPIPS $\downarrow$
& 0.351
& \second{0.335}
& \best{0.333}
& 0.306
& \second{0.298}
& \best{0.266}
\\

& CLIP-IQA $\uparrow$
& 0.540
& \second{0.544}
& \best{0.556}
& 0.612
& \second{0.640}
& \best{0.669}
\\

& NIQE $\downarrow$
& 3.648
& \second{3.514}
& \best{3.349}
& \second{3.366}
& 3.482
& \best{3.263}
\\

& MUSIQ $\uparrow$
& \second{60.82}
& \best{61.08}
& 60.55
& 65.42
& \second{66.49}
& \best{67.74}
\\

& Q-Align $\uparrow$
& 4.349
& \best{4.488}
& \second{4.400}
& \best{4.657}
& 4.603
& \second{4.617}
\\

\midrule

\multirow{4}{*}{\shortstack{\texttt{Real}\\SL}}
& CLIP-IQA $\uparrow$
& 0.507
& \second{0.523}
& \best{0.572}
& 0.566
& \best{0.604}
& \second{0.595}
\\

& NIQE $\downarrow$
& 4.616
& \second{4.516}
& \best{4.150}
& 4.671
& \second{4.462}
& \best{4.013}
\\

& MUSIQ $\uparrow$
& 57.62
& \second{58.26}
& \best{60.32}
& 62.35
& \second{63.81}
& \best{63.89}
\\

& Q-Align $\uparrow$
& 4.247
& \second{4.334}
& \best{4.354}
& 4.289
& \second{4.416}
& \best{4.451}
\\

\midrule

\multirow{4}{*}{\shortstack{\texttt{Real}\\MRL}}
& CLIP-IQA $\uparrow$
& \second{0.533}
& 0.515
& \best{0.536}
& 0.567
& \second{0.571}
& \best{0.633}
\\

& NIQE $\downarrow$
& 4.484
& \best{4.201}
& \second{4.363}
& 4.479
& \best{4.253}
& \second{4.341}
\\

& MUSIQ $\uparrow$
& 55.31
& \second{55.57}
& \best{56.30}
& 60.24
& \second{60.42}
& \best{62.87}
\\

& Q-Align $\uparrow$
& 3.875
& \second{3.907}
& \best{3.919}
& \best{4.065}
& 4.009
& \second{4.045}
\\

\bottomrule
\end{tabular}
\end{adjustbox}

\vspace{-2mm}
\end{table*}

%% file: Table/few_shot.tex
\begin{table}[h]
\centering
\caption{Few-shot adaptation with unpaired target observations on the \texttt{Screen} SL domain.}
\label{tab:fewshot_full}
\footnotesize
\setlength{\tabcolsep}{4.0pt}
\renewcommand{\arraystretch}{1.0}

\begin{tabular}{c|cccc}
\toprule
Target Images
& PSNR$\uparrow$
& LPIPS$\downarrow$
& MUSIQ$\uparrow$
& Q-Align$\uparrow$ \\
\midrule
0  & 15.075 & 0.332 & 53.651 & 3.725 \\
5  & 19.551 & 0.277 & 64.283 & 4.384 \\
15 & 19.654 & 0.271 & 64.579 & 4.420 \\
25 & 20.034 & 0.270 & 65.113 & 4.436 \\
50 & \textbf{20.462} & \textbf{0.266} & \textbf{65.303} & \textbf{4.473} \\
\bottomrule
\end{tabular}
\end{table}

%% file: Table/clear_image_collection_ablation.tex
\begin{table}[h]
\centering
\footnotesize
\setlength{\tabcolsep}{4pt}
\renewcommand{\arraystretch}{1.1}

\caption{
Sensitivity to the clear reference dataset on the \texttt{Real} SL domain.
}
\label{tab:clear_dataset_sensitivity}

\begin{tabular}{c|ccccc}
\toprule
$\mathcal{D}_{\mathrm{clr}}$
& CLIP-IQA$\uparrow$
& NIQE$\downarrow$
& MANIQA$\uparrow$
& MUSIQ$\uparrow$
& Q-Align$\uparrow$ \\
\midrule

Flickr2K
& 0.549
& 4.003
& 0.372
& 60.884
& 4.380
\\

DIV2K
& 0.572
& 4.150
& 0.367
& 60.320
& 4.354
\\

\bottomrule
\end{tabular}

\end{table}